\documentclass[aps,twocolumn,pra,superscriptaddress]{revtex4}
\usepackage{epsfig,graphicx,times}
\usepackage{amstext}
\usepackage{amsmath}
\usepackage{mathrsfs}
\usepackage{amssymb}
\usepackage{graphicx}
\usepackage{latexsym}
\usepackage{bm}
\usepackage{float}
\usepackage[colorlinks,citecolor=blue,linkcolor=blue,hyperindex,CJKbookmarks,driverfallback=dvipdfm]{hyperref}

\begin{document}

\title{Resonance-dominant optomechanical entanglement in open quantum systems}
\author{Cheng Shang \footnote{c-shang@iis.u-tokyo.ac.jp}}
\affiliation{Department of Physics, The University of Tokyo, 5-1-5 Kashiwanoha, Kashiwa, Chiba 277-8574, Japan}
\affiliation{Analytical quantum complexity RIKEN Hakubi Research Team, RIKEN Center for Quantum Computing (RQC), 2-1 Hirosawa, Wako, Saitama 351-0198, Japan}
\author{Hongchao Li \footnote{lhc@cat.phys.s.u-tokyo.ac.jp}}
\affiliation{Department of Physics, The University of Tokyo, 7-3-1 Hongo, Tokyo 113-0033, Japan}
\date{\today}

\begin{abstract}
Motivated by entanglement protection, our work utilizes a resonance effect to enhance optomechanical entanglement in the coherent-state representation. We propose a filtering model to filter out the significant detuning components between a thermal-mechanical mode and its surrounding heat baths in the weak coupling limit. We reveal that protecting continuous-variable entanglement involves the elimination of degrees of freedom associated with significant detuning components, thereby resisting decoherence. We construct a nonlinear Langevin equation of the filtering model and numerically show that the filtering model doubles the robustness of the stationary maximum optomechanical entanglement to the thermal fluctuation noise and mechanical damping. Furthermore, we generalize these results to an optical cavity array with one oscillating end-mirror to investigate the long-distance optimal optomechanical entanglement transfer. Our study breaks new ground for applying the resonance effect to protect quantum systems from decoherence and advancing the possibilities of large-scale quantum information processing and quantum network construction.
\end{abstract}

\maketitle

\section{Introduction}
Entanglement is an essential feature of quantum systems and one of the most striking phenomena of quantum theory~\cite{ref-1}, allowing for inseparable quantum correlations shared by distant parties~\cite{ref-2}. Entanglement is crucial in quantum information processing and network building~\cite{ref-3,ref-4,ref-5}. Studying entanglement properties from the perspectives of discrete and continuous variables is significant for further understanding the quantum-classical correspondence~\cite{ref-6,ref-7}. So far, the bipartite entanglement for a microscopic system of discrete variables with a few degrees of freedom has been studied in detail~\cite{ref-8}. A primary example of this is a two-qubit system. To quantify entanglement, concurrence~\cite{ref-9}, negativity~\cite{ref-10}, or the von Neumann entropy~\cite{ref-11} are frequently used in previous studies. 

Nevertheless, exploring bipartite entanglement in a macroscopic system of continuous variables with a large number of degrees of freedom has remained elusive~\cite{ref-12,ref-13,ref-14,ref-15}. Unfortunately, entanglement is fragile due to decoherence from inevitable dissipative couplings between an entangled system and its surrounding environment. Therefore, generating, measuring, and protecting entanglement in open quantum systems have raised widespread interest in various branches of physics and have been expected to be demonstrated to date~\cite{ref-16}.

Cavity optomechanical systems are based on couplings due to radiation pressure between electromagnetic and mechanical degrees of freedom~\cite{ref-17}. They provide a desirable mesoscopic platform for studying continuous-variable entanglement between optical cavity fields and macroscopic mechanical oscillators with vast degrees of freedom in open quantum systems~\cite{ref-18,ref-19}. Thanks to the rapid-developing field of microfabrication, quantum effects are becoming more significant as the size of devices is shrinking~\cite{ref-20,ref-21}. Remarkable progress has been made in generating entanglement by manipulating macroscopic nanomechanical oscillators with high precision~\cite{ref-22,ref-23}. Some landmark contributions have been achieved for an optomechanical entanglement measure~\cite{ref-24,ref-25}, such as using logarithmic negativity to calculate an upper bound of distillable optomechanical entanglement~\cite{ref-26}. 

Protecting the maximum optomechanical entanglement in open quantum systems has recently become a research focus. Many schemes have been proposed, such as protecting entanglement via synthetic magnetism in loop-coupled cavity optomechanical systems from thermal noise and dark mode~\cite{ref-27}, realizing phase-controlled asymmetric entanglement in cavity optomechanical systems of whispering-gallery-mode~\cite{ref-28}, achieving and preserving the optimal quality of nonreciprocal optomechanical entanglement via the Sagnac effect in a spinning cavity optomechanical systems evanescently coupled with a tapered fiber~\cite{ref-29,ref-30}, and via general dark-mode control to accomplish thermal-noise-resistant entanglement~\cite{ref-31}. 

However, the auxiliary protection of optomechanical entanglement in these schemes all work in hybrid cavity optomechanical systems, which inevitably brings about trilateral and even multilateral entanglement problems~\cite{ref-32}, such as photon-phonon-atom entanglement~\cite{ref-33}. In this sense, it is essential to develop methods of protecting the intrinsic bilateral optomechanical entanglement in hybrid cavity optomechanical systems from potential interference caused by additional types of degrees of freedom~\cite{ref-34}. With this motivation, we aim to protect a prototypical optomechanical entanglement in cavity optomechanical systems.

Currently, intriguing schemes have been proposed to achieve the frequency resonance of the system by using laser driving, thereby protecting bilateral mechanical entanglement in doubly resonant cavity optomechanical systems~\cite{ref-35,ref-36} and photon-atom entanglement in the Rabi model~\cite{ref-37}. Inspired by this, we propose to utilize the high-frequency resonance effect in a Fabry-P\'{e}rot cavity to protect the maximal value of optomechanical entanglement. In the weak coupling limit, a clear-cut physical mechanism is employed to reduce Brownian noise and dissipation, which involves filtering out components with significant mismatched coupling frequencies between a mechanical mode and its thermal reservoir by leveraging the high-frequency resonance effect. The present theoretical conjecture can be materialized in an experiment by laser-driving the optical cavity field to resonate with a high-frequency and high-quality-factor mechanical resonator coupled to a Markovian structured environment. We can observe resonance-dominant optomechanical entanglement using a homodyne detection scheme~\cite{ref-38,ref-39} or a cavity-assisted measurement scheme~\cite{ref-40,ref-41}.

To attain our goal, we start by constructing the Hamiltonian of the cavity optomechanical system under the coherent-state representation. We then derive its associated nonlinear Langevin equations, which are consistent with the results in Ref.~\cite{ref-24} but originate from the coherent-state representation. We finally propose a theory of resonance-dominant optomechanical entanglement in continuous-variable systems. When the mechanical mode and surrounding heat baths satisfy the conditions of weak coupling and high-frequency resonance, we point out that the filtering model protects the stationary maximum optomechanical entanglement. In particular, we quantitatively observe that a resonance effect doubles the robustness of the mechanical damping and thermal fluctuation noise from the environment and reveals its physical reason. This result first unveils a hitherto overlooked aspect of applying a resonance effect to entanglement protection. We further extend these results to an array of optical cavities with one oscillating end-mirror and investigate the remote optomechanical entanglement, which helps achieve optimal optomechanical entanglement transmission for quantum information processing.

The remainder of this paper is organized as follows. In Sec.~{\ref{section2}}, we construct the Hamiltonian of the physical system and reproduce the results of nonlinear Langevin equations in Ref.~\cite{ref-24} in the coherent-state representation. In Sec.~{\ref{section3}}, we propose a theory of resonance-dominant optomechanical entanglement in continuous-variable systems and show the results for the maximum optomechanical entanglement protection. In addition, we present a potential experimental implementation of this scheme. In Sec.~{\ref{section4}}, we extend these findings to an array of optical cavities with one oscillating end-mirror, investigating the remote optimal optomechanical entanglement transmission for application purposes. Finally, in Sec.~{\ref{section5}}, we summarize our findings and discuss the outlook for future research.

\section{Reformulating Dynamics in Coherent State Representation}\label{section2}
\subsection{Construction of Hamiltonian}

\begin{figure}[h]
\centering
\includegraphics[angle=0,width=0.46\textwidth]{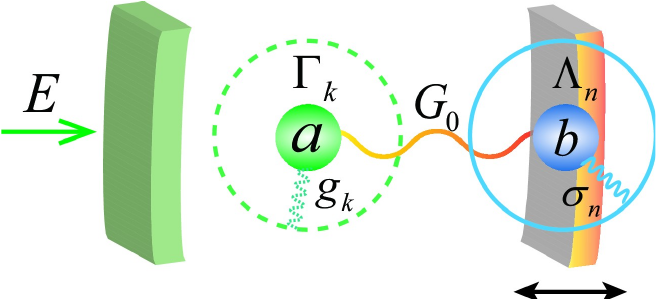}
\caption{A cavity optomechanical system driven by a monochromatic laser. The optical and mechanical modes are coupled via radiation pressure while independently coupled to their respective reservoirs.}\label{Fig-0}
\end{figure}

We first construct an open-quantum-system description of a cavity optomechanical system in the coherent-state representation as shown in Fig.~\ref{Fig-0}. The Fabry-P\'{e}rot cavity, known as the simplest optical resonator structure, is additionally driven by a monochromatic laser, described by the radiation-pressure interaction between an optical cavity field and a vibrating end mirror, which applies to a wide variety of optomechanical devices, including microwave resonators~\cite{ref-43}, optomechanical crystals~\cite{ref-44}, and setups with the membrane inside a cavity~\cite{ref-45}. 

Meanwhile, we assume that a cavity optomechanical system is coupled to two reservoirs. The optical mode is coupled to a reservoir characterized by zero-temperature electromagnetic modes, while the mechanical mode is coupled to another reservoir consisting of harmonic oscillators at thermal equilibrium~\cite{ref-46}. In the Heisenberg picture, the system and environment evolve in time under the influence of the total Hamiltonian that reads
\begin{eqnarray}
{H_{\rm{T}}} = {H_{\rm{S}}} + {H_{\rm{B}}},\label{eq-1}
\end{eqnarray}
where
\begin{eqnarray}
{H_{\rm{S}}} \!\!&=&\!\!  + \hbar {\Delta _0}{a^\dag }a + \hbar {\omega _{\rm{m}}}{b^\dag }b - \hbar \frac{{{G_0}}}{{\sqrt 2 }}{a^\dag }a\left( {{b^\dag } + b} \right)\nonumber\\
 &&+ i\hbar \left( {E{a^\dag } - {E^*}a} \right),\label{eq-2}\\
{H_{\rm{E}}} \!\!&=&\!\!  + \hbar \sum\limits_k {{\omega _k}\Gamma _k^\dag } {\Gamma _k} + \hbar \sum\limits_k {{g_k}\left( {\Gamma _k^\dag a + {a^\dag }{\Gamma _k}} \right)} \label{eq-3}\\
 &&+ \hbar \sum\limits_n {{\omega _n}\Lambda _n^\dag } {\Lambda _n} - i\hbar \sum\limits_n {\frac{{{\sigma _n}}}{2}\left( {\Lambda _n^\dag  - {\Lambda _n}} \right)} \left( {{b^\dag } + b} \right), \nonumber
\end{eqnarray}
with ${a^\dag }$ $\left( a \right)$ denoting ${b^\dag }$ $\left( b \right)$ are the creation (annihilation) operators of the optical mode and the mechanical mode, respectively. Laser detuning from the cavity resonance is ${\Delta _0} = {\omega _{\rm{c}}} - {\omega _L}$, where ${\omega _{\rm{c}}}$ is the cavity characteristic frequency and ${\omega _L}$ the is driving laser frequency. The characteristic frequency and effective mass of the mechanical oscillator are ${\omega _{\rm{m}}}$ and $m$, respectively. The optomechanical coupling coefficient is ${G_0} = \left( {{{{\omega _{\rm{c}}}} \mathord{/{\vphantom {{{\omega _{\rm{c}}}} L}} \kern-\nulldelimiterspace} L}} \right)\sqrt {{\hbar  \mathord{/{\vphantom {\hbar  {m{\omega _{\rm{m}}}}}} \kern-\nulldelimiterspace} {m{\omega _{\rm{m}}}}}}$, with $L$ being the cavity length. The complex amplitude of the driving laser is $E$. In addition, $\Gamma _{k}^\dag $ $\left( {{\Gamma _{k}}} \right)$ and $\Lambda _{n}^\dag $ $\left( {{\Lambda _{n}}} \right)$ for ${k} \in \left\{ {1,2,3 \cdot  \cdot  \cdot, + \infty } \right\}$ and ${n} \in \left\{ {1,2,3 \cdot  \cdot  \cdot, + \infty } \right\}$ are, respectively, the creation (annihilation) operators of the reservoirs for the optical mode and the mechanical mode. The harmonic-oscillator reservoirs have closely spaced frequencies corresponding to photons and phonons, denoted by ${\omega _{k}}$ and ${{\omega _{n}}}$, respectively. The real numbers ${g_{k}}$ and ${\sigma _{n}}$ represent the coupling strengths between the subsystem and the $n$th reservoir mode, respectively. Details of the derivations of the total Hamiltonian~(\ref{eq-1}) are attached in Appendix~\ref{appendix-A} \cite{ref-47}.

\subsection{Nonlinear Langevin equations}
A reasonable description of the dynamics in an open quantum system should include photon losses in the optical cavity field and the Brownian noise acting on the vibrating end mirror. By substituting Eq.~(\ref{eq-1}) into the Heisenberg equation and taking into account the dissipation and noise terms, we obtain a set of closed integrodifferential equations (see Appendix \ref{appendix-B} for the derivation~\cite{ref-47}) for the operators of the optical mode and mechanical mode as follows:
\begin{eqnarray}
\dot q &=& {\omega _{\rm{m}}}p,\label{eq-4}\\
\dot p &=&  - {\omega _{\rm{m}}}q - {\gamma _{\rm{m}}}p + {G_0}{a^\dag }a + \xi,\label{eq-5}\\
\dot a &=&  - \left( {\kappa  + i{\Delta _0}} \right)a + i{G_0}aq + E + \sqrt {2\kappa } {a_{{\rm{in}}}},\label{eq-6}
\end{eqnarray}
where $q = {{\left( {{b^\dag } + b} \right)} \mathord{/{\vphantom {{\left( {{b^\dag } + b} \right)} {\sqrt 2 }}} \kern-\nulldelimiterspace} {\sqrt 2 }}$ and $p = {{i\left( {{b^\dag } - b} \right)} \mathord{/{\vphantom {{i\left( {{b^\dag } - b} \right)} {\sqrt 2 }}} \kern-\nulldelimiterspace} {\sqrt 2 }}$ are the dimensionless position and momentum operators of the vibrating end mirror. We assume that the decay rate of the optical cavity is $\kappa $ and set the mechanical damping rate as ${\gamma _{\rm{m}}} = {\omega _{\rm{m}}}\gamma $. The dissipative terms $\kappa $ and $\gamma $ are proportional to the square of the coupling strength between the subsystem and the reservoir ${g_{k}}$ and ${\sigma _{n}}$, respectively. The optical Langevin force ${a_{{\rm{in}}}}$ represents the field incident to the cavity and is assumed to be in the vacuum state. Its specific expression and the correlation function~\cite{ref-48} are
\begin{eqnarray}
{a_{{\rm{in}}}}\left( t \right) \!&=&\! \frac{{ - i}}{{\sqrt {2\pi } }}\sum\limits_{k} {{g_{k}}} \Gamma \left( {{t_0}} \right){e^{ - i{\omega _{k}}\left( {t - {t_0}} \right)}}, \nonumber\\ \langle {{a_{{\rm{in}}}}\left( t \right)a_{{\rm{in}}}^\dag \left( {t'} \right)} \rangle  \!&=&\! \delta \left( {t - t'} \right), \label{eq-7}
\end{eqnarray}
where ${t_0}$ represents the initial time. This correlation function is true for optical fields at room temperature or microwaves at a cryostat. 

In contrast, the Brownian noise operator is given by
\begin{eqnarray}
\xi \left( t \right) \!=\! \sum\limits_n \!{\frac{{i{\sigma _n}}}{{\sqrt 2 }}}\! \left[ {\Lambda _n^\dag \left( {{t_0}} \right){e^{i{\omega _n}\left( {t - {t_0}} \right)}} \!-\! {\Lambda _n}\left( {{t_0}} \right){e^{ - i{\omega _n}\left( {t - {t_0}} \right)}}} \right], \label{eq-8}
\end{eqnarray}
The mechanical damping force $\xi $ is non-Markovian in general~\cite{ref-49}, but it can be treated as Markovian if the following two conditions are met: the thermal bath occupation number satisfies ${{\bar n}} \gg {\rm{1}}$; the mechanical quality factor satisfies $Q = {{{\omega _{\rm{m}}}} \mathord{\left/{\vphantom {{{\omega _{\rm{m}}}} {{\gamma _{\rm{m}}} = {1 \mathord{\left/{\vphantom {1 \gamma }} \right.\kern-\nulldelimiterspace} \gamma }}}} \right.\kern-\nulldelimiterspace} {{\gamma _{\rm{m}}} = {1 \mathord{\left/{\vphantom {1 \gamma }} \right.\kern-\nulldelimiterspace} \gamma }}} \gg 1$. These conditions are well satisfied in the majority of contemporary experimental setups, which validates the use of the standard Markovian delta-correlation~\cite{ref-50,ref-51}:
\begin{eqnarray}
{{\left\langle {\xi \left( t \right)\xi \left( {t'} \right) + \xi \left( {t'} \right)\xi \left( t \right)} \right\rangle } \mathord{/{\vphantom {{\left\langle {\xi \left( t \right)\xi \left( {t'} \right) + \xi \left( t \right)\xi \left( {t'} \right)} \right\rangle } 2}}\kern-\nulldelimiterspace} 2} \approx {\gamma _{\rm{m}}}\left( {2{\bar n} + 1} \right)\delta \left( {t - t'} \right), \label{eq-9}
\end{eqnarray}
where ${{\bar n}} = {\left[ {\exp \left( {{{\hbar {\omega _{\rm{m}}}} \mathord{\left/{\vphantom {{\hbar {\omega _{\rm{m}}}} {{k_{\rm{B}}}T}}} \right.\kern-\nulldelimiterspace} {{k_{\rm{B}}}T}}} \right) - 1} \right]^{ - 1}}$ is the mean thermal excitation number with the Boltzmann constant ${k_{\rm{B}}}$ and the end-mirror temperature $T$.

So far, we have constructed the total Hamiltonian of the optomechanical system under the coherent-state representation and completely reproduced the results of the nonlinear Langevin equations in Ref.~\cite{ref-24}, which provides solid support for the filtering model dominated by the resonance effect discussed later. We stress that deriving the Langevin equation from the total Hamiltonian provides a clear picture in explicitly revealing the specific form of the interaction between the system and the environment and the physical origin of each term in nonlinear Langevin equations, in comparison to the implicit treatment of such interactions in the Lindblad master equation.

\section{Resonance-dominant optomechanical entanglement}\label{section3}
\subsection{Filtering Model}
In the preceding section, the Hamiltonian~(\ref{eq-1}) describes an original interaction between an optomechanical system and its surrounding environment. This section proposes a resonant filtering model in the weak coupling limit between the system and the heat bath. It uses a high-frequency resonance between the mechanical mode and its thermal reservoirs to filter out non-resonant degrees of freedom and achieve quantum coherence protection.

To discuss the frequency relation between the mechanical mode and its thermal reservoirs, we introduce the frequency transformation $\tilde b\left( t \right) = b\left( t \right)\exp \left( {i{\omega _{\rm{m}}}t} \right)$ and ${\tilde \Lambda _n}\left( t \right) = {\Lambda _n}\left( t \right)\exp \left( {i{\omega _n}t} \right)$ for ${b}\left( t \right)$ and ${\Lambda _{{n}}}\left( t \right)$~\cite{ref-46} in the interaction picture. After that, the Hamiltonian~(\ref{eq-1}) reads
\begin{eqnarray}
{H_{\rm{T}}} =  \!\!\!\!\!&&+ \hbar {\Delta _0}{a^\dag }a + \hbar {\omega _{\rm{m}}}{b^\dag }b + i\hbar \left( {E{a^\dag } - {E^*}a} \right) \nonumber\\
 &&- \hbar \frac{{{G_0}}}{{\sqrt 2 }}{a^\dag }a\left( {{{\tilde b}^\dag }{e^{i{\omega _{\rm{m}}}t}} + \tilde b{e^{ - i{\omega _{\rm{m}}}t}}} \right) + \hbar \sum\limits_k {{\omega _k}\Gamma _k^\dag } {\Gamma _k} \nonumber\\
 &&+ \hbar \sum\limits_k {{g_k}\left( {\Gamma _k^\dag a + {a^\dag }{\Gamma _k}} \right)}  + \hbar \sum\limits_n {{\omega _n}\Lambda _n^\dag } {\Lambda _n} \label{eq-10}\\
 &&- i\hbar \sum\limits_n {\frac{{{\sigma _n}}}{2}\left[ {\tilde \Lambda _n^\dag \tilde b{e^{i\left( {{\omega _n} - {\omega _{\rm{m}}}} \right)t}} - {{\tilde b}^\dag }{{\tilde \Lambda }_n}{e^{ - i\left( {{\omega _n} - {\omega _{\rm{m}}}} \right)t}}} \right]} \nonumber\\
 &&- i\hbar \sum\limits_n {\frac{{{\sigma _n}}}{2}} \left[ {\tilde \Lambda _n^\dag {{\tilde b}^\dag }{e^{i\left( {{\omega _n} + {\omega _{\rm{m}}}} \right)t}} - \tilde b{{\tilde \Lambda }_n}{e^{ - i\left( {{\omega _n} + {\omega _{\rm{m}}}} \right)t}}} \right].\nonumber
\end{eqnarray}
\begin{figure*}[t]
\centering
\includegraphics[angle=0,width=0.92\textwidth]{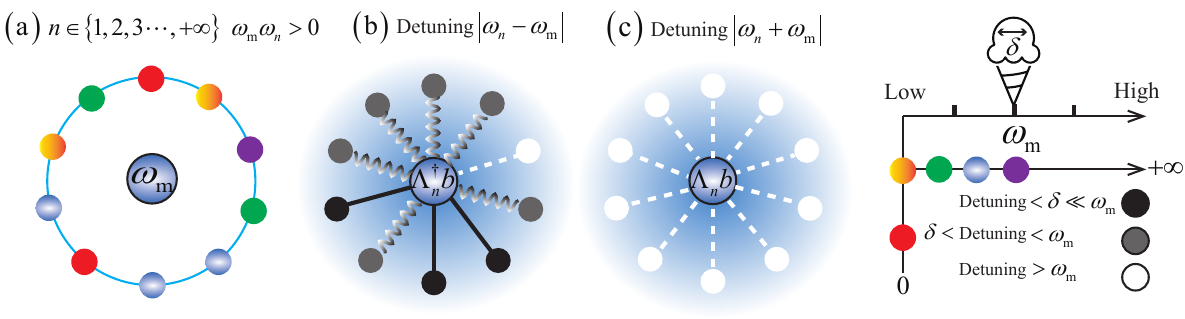}
\caption{Schematic diagram of the filtering model. $\left( {\rm{a}} \right)$ The coupling between a high-frequency mechanical oscillator and co-directional thermal reservoirs. The frequency of the thermal reservoirs sequentially transits from zero to positive infinity in rainbow color order. In $\left( {\rm{b}} \right)$, the terms of $\tilde \Lambda _{{n}}^\dag \tilde b$ and ${{\tilde b}^\dag }{{\tilde \Lambda }_{{n}}}$ show the high-frequency resonance effect, while in $\left( {\rm{c}} \right)$, the terms of $\tilde \Lambda _{{n}}^\dag {{\tilde b}^\dag }$ and $\tilde b{{\tilde \Lambda }_{{n}}}$ exhibit large detuning effects. The black, gray, and white colors correspond to the high-frequency resonance, moderately detuned, and highly detuned modes of the heat bath compared to the frequency of the single-mode mechanical oscillator ${\omega _{\rm{m}}}$. The parameters $\delta $ and ${\omega _{\rm{m}}}$ can be modulated by coherent laser driving~\cite{ref-53,ref-54}.}\label{Fig-1}
\end{figure*}

As aforementioned, our physical model describes a Markovian process in the weak coupling limit $\gamma  \ll 1$, which corresponds to Eq.~(\ref{eq-10}) satisfying the weak-coupling limit ${\sigma _{{n}}} \ll 1$ for ${{n}} \in \left\{ {1,2,3 \cdot  \cdot  \cdot, + \infty } \right\}$~\cite{ref-52}. See Fig.~\ref{Fig-1} for a schematic diagram, according to the rotating-wave approximation, we eliminate the fast-oscillating terms from Eq.~(\ref{eq-10}), and then recovering $b\left( t \right) = \tilde b\left( t \right){e^{ - i{\omega _{\rm{m}}}t}}$ and ${\Lambda _{{n}}}\left( t \right) = {{\tilde \Lambda }_{{n}}}\left( t \right){e^{ - i{\omega _{{n}}}t}}$, we classify the filtering model reduced from Eq.~(\ref{eq-10}) as follows.

The high-frequency resonance region is defined by ${\omega _{\rm{m}}}{\omega _{{n}}} > 0$ and ${\omega _{{n}}} \in \left( {0, + \infty } \right)$. We here propose to filter out the strongly non-resonant contributions ${\tilde \Lambda _{{n}}^\dag {{\tilde b}^\dag }}$ and ${\tilde b{{\tilde \Lambda }_{{n}}}}$ mechanically; see Sec.~\ref{section3}-D for possible experimental realizations. Keeping only the resonant terms ${\tilde \Lambda _{{n}}^\dag \tilde b}$ and ${{{\tilde b}^\dag }{{\tilde \Lambda }_{{n}}}}$, the filtering model is
\begin{eqnarray}
{H_{\rm{F}}} = \!\!\!\!\!&&+ \hbar {\Delta _0}{a^\dag }a + \hbar {\omega _{\rm{m}}}{b^\dag }b + i\hbar \left( {E{a^\dag } - {E^*}a} \right)\nonumber\\
 &&- \hbar \frac{{{G_0}}}{{\sqrt 2 }}{a^\dag }a\left( {b + {b^\dag }} \right) + \hbar \sum\limits_k {{\omega _k}\Gamma _k^\dag } {\Gamma _k}\nonumber\\
 &&+ \hbar \sum\limits_k {{g_k}\left( {\Gamma _k^\dag a + {a^\dag }{\Gamma _k}} \right)}  + \hbar \sum\limits_n {{\omega _n}\Lambda _n^\dag } {\Lambda _n}\nonumber\\
 &&- i\hbar \sum\limits_n {\frac{{{\sigma _n}}}{2}\left[ {\Lambda _n^\dag b - {b^\dag }{\Lambda _n}} \right]} \label{eq-11}.
\end{eqnarray}
The resonance terms ${\tilde \Lambda _{{n}}^\dag \tilde b}$ and ${{{\tilde b}^\dag }{{\tilde \Lambda }_{{n}}}}$ in this region describe the exchange of quanta between the mechanical mode and its $n$th thermal reservoir mode \cite{ref-18}. In contrast, the high-frequency inverse-resonance region is defined by ${\omega _{\rm{m}}}{\omega _{n}} < 0$ and ${\omega _{{n}}} \in \left( {0, - \infty } \right)$. Keeping only the terms of ${\tilde \Lambda _{{n}}^\dag {{\tilde b}^\dag }}$ and ${\tilde b{{\tilde \Lambda }_{{n}}}}$, the inverse-filtering model reads
\begin{eqnarray}
H_{\rm{F}}^{\rm{I}} =  \!\!\!\!\!&&+ \hbar {\Delta _0}{a^\dag }a + \hbar {\omega _{\rm{m}}}{b^\dag }b + i\hbar \left( {E{a^\dag } - {E^*}a} \right)\nonumber\\
 &&- \hbar \frac{{{G_0}}}{{\sqrt 2 }}{a^\dag }a\left( {b + {b^\dag }} \right) + \hbar \sum\limits_k {{\omega _k}\Gamma _k^\dag } {\Gamma _k}\nonumber\\
 &&+ \hbar \sum\limits_k {{g_k}\left( {\Gamma _k^\dag a + {a^\dag }{\Gamma _k}} \right)}  + \hbar \sum\limits_n {{\omega _n}\Lambda _n^\dag } {\Lambda _n}\nonumber\\
 &&- i\hbar \sum\limits_n {\frac{{{\sigma _n}}}{2}\left[ {\Lambda _n^\dag {b^\dag } - b{\Lambda _n}} \right]}.\label{eq-12}
\end{eqnarray}
The inverse-resonance terms ${\tilde \Lambda _{{n}}^\dag {{\tilde b}^\dag }}$ and ${\tilde b{{\tilde \Lambda }_{{n}}}}$ in this region represent a two-mode squeezing interaction between the mechanical mode and its $n$th thermal reservoir mode, and the parametric amplification relies on the two-mode squeezing interaction \cite{ref-55}.

\subsection{The Lyapunov equation for the steady-state correlation matrix}
In order to comprehend the impact of resonance effects between a mechanical mode and its thermal reservoirs on the strength of an optomechanical system, it is crucial to gain insight into the structure of optomechanical correlation in open quantum systems. For this purpose, we use the Lyapunov equation to compute the steady-state correlation matrix between subsystems and obtain the optomechanical entanglement strength \cite{ref-56}. Without loss of generality, we take the high-frequency resonance regime as an example of deriving the Lyapunov equation in terms of the steady-state correlation matrix.

By deriving the Heisenberg equation of motion of the resonant Hamiltonian ${H_{\rm{F}}}$~(\ref{eq-11}), we obtain nonlinear Langevin equations that govern the dynamical behavior of the optomechanical system in the high-frequency resonance regime. The nonlinear Langevin equations are written as (see Appendix \ref{appendix-C} for details)
\begin{eqnarray}
\dot q &=& {\omega _{\rm{m}}}p + \frac{\gamma }{4}\dot p + \frac{1}{2}\xi ',\label{eq-13}\\
\dot p &=&  - {\omega _{\rm{m}}}q - \frac{\gamma }{4}\dot q + {G_0}{a^\dag }a + \frac{1}{2}\xi ,\label{eq-14}\\
\dot a &=&  - \left( {\kappa  + i{\Delta _0}} \right)a + i{G_0}aq + E + \sqrt {2\kappa } {a_{{\rm{in}}}},\label{eq-15}
\end{eqnarray}
where the Brownian noise operator reads
\begin{eqnarray}
\xi '\!\left( t \right) \!=\!\! \sum\limits_n \! {\frac{{{\sigma _n}}}{{\sqrt 2 }}} \!\!\left[ {\Lambda _n^\dag \left( {{t_0}} \right){e^{i{\omega _n}\left( {t - {t_0}} \right)}} \!\!+\! {\Lambda _n}\left( {{t_0}} \right){e^{ - i{\omega _n}\left( {t - {t_0}} \right)}}} \right]\!\!.\label{eq-16}
\end{eqnarray}
In the weak-coupling limit $\gamma  \ll 1$, by substituting Eqs.~(\ref{eq-13}) and (\ref{eq-14}) into each other and neglecting small terms in Eqs.~(\ref{eq-13})-(\ref{eq-14}), we obtain the reduced equations
\begin{eqnarray}
\dot q &=& {\omega _{\rm{m}}}p - \frac{{{\gamma _{\rm{m}}}}}{4}q + \frac{1}{2}\xi '\label{eq-17}\\
\dot p &=&  - {\omega _{\rm{m}}}q - \frac{{{\gamma _{\rm{m}}}}}{4}p + {G_0}{a^\dag }a + \frac{1}{2}\xi. \label{eq-18}
\end{eqnarray}
In the weak-coupling limit $\gamma  \ll 1$, the Brownian noise operator $\xi '\left( t \right)$ has the same delta-correlated form as $\xi \left( t \right)$.

The nonlinear Langevin equations~(\ref{eq-15}), (\ref{eq-17}) and (\ref{eq-18}) are inherently nonlinear as they contain a product of the photon operator and dimensionless position operator of the mechanical phonon, $aq$, as well as a quadratic term in photon operators, ${a^\dag }a$. Using the standard mean-field method~\cite{ref-57} to solve Eqs.~(\ref{eq-15}), (\ref{eq-17}) and (\ref{eq-18}), we start by splitting each Heisenberg operator into the classical mean values and quantum fluctuation operators, i.e., $a = {\alpha _{\rm{s}}} + \delta a$ as in ${a^\dag } = \alpha _{\rm{s}}^* + \delta {a^\dag }$, $q = {q_{\rm{s}}} + \delta q$, and $p = {p_{\rm{s}}} + \delta p$, thereby linearizing these equations. Adopting the above approach and inserting these expressions into nonlinear Langevin equations~(\ref{eq-15}), (\ref{eq-17}) and (\ref{eq-18}), we find the solution of the mean values for the classical steady state given by ${p_{\rm{s}}} = {{\gamma {q_{\rm{s}}}} \mathord{/{\vphantom {{\gamma {q_{\rm{s}}}} 4}} \kern-\nulldelimiterspace} 4} \approx 0$, ${q_{\rm{s}}} = {{{G_0}\alpha _{\rm{s}}^*{\alpha _{\rm{s}}}} \mathord{/{\vphantom {{{G_0}\alpha _{\rm{s}}^*{\alpha _{\rm{s}}}} {{\omega _{\rm{m}}}}}} \kern-\nulldelimiterspace} {{\omega _{\rm{m}}}}}$, and ${\alpha _{\rm{s}}} = {E \mathord{/{\vphantom {E {\left( {\kappa  + i\Delta } \right)}}}  \kern-\nulldelimiterspace} {\left( {\kappa  + i\Delta } \right)}}$, where we set normalization of the detuning frequency of the optical field as $\Delta  = {\Delta _0} - {{G_0^2\alpha _{\rm{s}}^*{\alpha _{\rm{s}}}} \mathord{/{\vphantom {{iG_0^2\alpha _{\rm{s}}^*{\alpha _{\rm{s}}}} {{\omega _{\rm{m}}}}}} \kern-\nulldelimiterspace} {{\omega _{\rm{m}}}}}{\rm{ }}$. The parameter regime for generating optomechanical entanglement is the one with a large amplitude of the driving laser $E$, i.e., ${\alpha _{\rm{s}}} \gg \delta a$ and $\alpha _{\rm{s}}^* \gg \delta {a^\dag }$. By dropping the contribution of terms of second orders in quantum fluctuations $\delta a\delta q$ and $\delta {a^\dag }\delta a$, we obtain the linearized Langevin equations
 \begin{eqnarray}
\delta \dot q \! &=& \! {\omega _{\rm{m}}}\delta p - \frac{{{\gamma _{\rm{m}}}}}{4}\delta q + \frac{1}{2}\xi ',\label{eq-19}\\
\delta \dot p \!&=&\!  - {\omega _{\rm{m}}}\delta q - \frac{{{\gamma _{\rm{m}}}}}{4}\delta p + {G_0}\left( {\alpha _{\rm{s}}^*\delta a + {\alpha _{\rm{s}}}\delta {a^\dag }} \right) + \frac{1}{2}\xi ,\label{eq-20}\\
\delta \dot a \!&=&\!  - \left( {\kappa  + i\Delta } \right)\delta a + i{G_0}{\alpha _{\rm{s}}}\delta q + \sqrt {2\kappa } {a_{{\rm{in}}}}. \label{eq-21}
\end{eqnarray}
By assuming the driving laser amplitude $E = \left| E \right|\exp \left( {i\varphi } \right)$, where $\left| E \right|$ is related to the input laser power $P$ by $\left| E \right| = \sqrt {{{2P\kappa } \mathord{\left/{\vphantom {{2P\kappa } {\hbar {\omega _L}}}} \right.\kern-\nulldelimiterspace} {\hbar {\omega _L}}}} $ and $\varphi $ denotes the phase of the laser field coupling to the optical cavity field, we choose $\varphi $ to satisfy $\tan \left( \varphi  \right) = {\Delta  \mathord{/{\vphantom {\Delta  \kappa }} \kern-\nulldelimiterspace} \kappa }$ so that ${{\alpha _{\rm{s}}}}$ may be real. 

The quadratures play an essential role in studying entanglement because they are used to quantify the correlations between different modes. We define the cavity field quadratures $\delta X = {{\left( {\delta a + \delta {a^\dag }} \right)} \mathord{/{\vphantom {{\left( {\delta a + \delta {a^\dag }} \right)} {\sqrt 2 }}} \kern-\nulldelimiterspace} {\sqrt 2 }}$ and $\delta Y = i{{\left( {\delta {a^\dag } - \delta a} \right)} \mathord{/{\vphantom {{\left( {\delta {a^\dag } - \delta a} \right)} {\sqrt 2 }}} \kern-\nulldelimiterspace} {\sqrt 2 }}$ as two observables that describe the quantum state of a cavity field mode, which can be measured using homodyne detection techniques. Accordingly, we define the orthogonal input noise operators ${X_{{\rm{in}}}} = {{( {\delta a_{{\rm{in}}}^\dag  + \delta {a_{{\rm{in}}}}} )} \mathord{/{\vphantom {{\left( {\delta a_{{\rm{in}}}^\dag  + \delta {a_{{\rm{in}}}}} \right)} {\sqrt 2 }}} \kern-\nulldelimiterspace} {\sqrt 2 }}$ and ${Y_{{\rm{in}}}} = i{{( {\delta a_{{\rm{in}}}^\dag  - \delta {a_{{\rm{in}}}}} )} \mathord{/{\vphantom {{\left( {\delta a_{{\rm{in}}}^\dag  - \delta {a_{{\rm{in}}}}} \right)} {\sqrt 2 }}} \kern-\nulldelimiterspace} {\sqrt 2 }}$, and thereby rewrite Eqs.~(\ref{eq-19})-(\ref{eq-21}) as
\begin{eqnarray}
\delta \dot q &=& {\omega _{\rm{m}}}\delta p - \frac{{{\gamma _{\rm{m}}}}}{4}\delta q + \frac{1}{2}\xi ',\label{eq-22}\\
\delta \dot p &=&  - {\omega _{\rm{m}}}\delta q - \frac{{{\gamma _{\rm{m}}}}}{4}\delta p + G\delta X + \frac{1}{2}\xi,\label{eq-23}\\
\delta \dot X &=&  - {\kappa}\delta X + \Delta \delta Y + \sqrt {2{\kappa}} {X_{{\rm{in}}}},\label{eq-24}\\
\delta \dot Y &=&  - {\kappa}\delta Y - \Delta \delta X + G\delta q + \sqrt {2{\kappa}} Y_{{\rm{in}}}, \label{eq-25}
\end{eqnarray}
where the effective optomechanical coupling is given by $G = \sqrt 2 {\alpha _{\rm{s}}}{G_0}$. 

For convenience, we concisely express a linearized Langevin Eqs.~(\ref{eq-22})-(\ref{eq-25}) for orthogonal operators in a matrix form,
\begin{eqnarray}
\dot \mu \left( t \right) = {{A}}\mu \left( t \right) + n\left( t \right), \label{eq-26}
\end{eqnarray}
where the component of each matrix is as follows: the transposes of the column vector of continuous variables fluctuation operators are written as ${\mu ^T}\left( t \right) = \left[ {\delta q\left( t \right),\delta p\left( t \right),\delta X\left( t \right),\delta Y\left( t \right)} \right]$; the transposes of the column vector of noise operators are denoted by ${n^T}\left( t \right) = \left[ {0.5\xi '\left( t \right),0.5\xi \left( t \right),\sqrt {2\kappa } {X_{{\rm{in}}}}\left( t \right),\sqrt {2\kappa } {Y_{{\rm{in}}}}\left( t \right)} \right]$; the coefficient matrix ${{A}}$ in terms of system parameters takes the form
\begin{eqnarray}
{{A}} = \left( {\begin{array}{*{20}{c}}
{ - 0.25{\gamma _{\rm{m}}}}&{{\omega _{\rm{m}}}}&0&0\\
{ - {\omega _{\rm{m}}}}&{ - 0.25{\gamma _{\rm{m}}}}&G&0\\
0&0&{ - \kappa }&\Delta \\
G&0&{ - \Delta }&{ - \kappa }
\end{array}} \right).\label{eq-27}
\end{eqnarray}
The solution of Eq.~(\ref{eq-26}) can be expressed as
\begin{eqnarray}
\mu \left( t \right) = {{M}}\left( t \right) \mu \left( {t_0} \right) + \int_{t_0}^t {{{M}}\left( \tau  \right)} n\left( {t - \tau } \right)d\tau,  \label{eq-28}
\end{eqnarray}
where ${M}$ is the matrix exponential ${{M}}\left( t \right) = \exp \left( {{\rm{A}}t} \right)$ and we assume the initial time as ${t_0} = 0$. The system is stable if and only if the real parts of all the eigenvalues of the matrix ${{A}}$ are negative. The eigenvalue equation det$\left| {{{A}} - \lambda {{\rm{I}}_4}} \right| = [ {{{\left( {0.25{\gamma _{\rm{m}}} + \lambda } \right)}^2} + \omega _{\rm{m}}^2} ][ {{{\left( {\kappa  + \lambda } \right)}^2} + {\Delta ^2}} ] - {\omega _{\rm{m}}}{G^2}\Delta  = 0$, where ${{{\rm{I}}_4}}$ denotes the four-dimensional identity matrix, can be reduced to the fourth-order equation ${C_0}{\lambda ^4} + {C_1}{\lambda ^3} + {C_2}{\lambda ^2} + {C_3}\lambda  + {C_4}{\rm{ = }}0$. The stability conditions can be derived by applying the Routh-Hurwitz criterion~\cite{ref-58} as follows: \!\!\!${C_0} >\!\! 0, \ {C_1} > 0, \ {C_1}{C_2} - {C_0}{C_3} > 0, \ \left( {{C_1}{C_2} - {C_0}{C_3}} \right){C_3} - C_1^2{C_4} > 0, \ {C_4} > 0$, yielding the following two nontrivial conditions: $\left( {\omega _{\rm{m}}^2 + {{\gamma _{\rm{m}}^2} \mathord{\left/
 {\vphantom {{\gamma _{\rm{m}}^2} {16}}} \right.
 \kern-\nulldelimiterspace} {16}}} \right)\left( {{\Delta ^2} + {\kappa ^2}} \right) - {\omega _{\rm{m}}}{G^2}\Delta  > 0$ and 
\begin{widetext}
\begin{eqnarray}
+{\gamma _{\rm{m}}}\kappa \left\{ {{\Delta ^4} + {\Delta ^2}\left( {\frac{{\gamma _{\rm{m}}^2}}{8} + {\gamma _{\rm{m}}}\kappa  + 2{\kappa ^2} - 2\omega _{\rm{m}}^2} \right) + \frac{1}{{256}}{{\left[ {16\omega _{\rm{m}}^2 + {{\left( {{\gamma _{\rm{m}}} + 4\kappa } \right)}^2}} \right]}^2}} \right\} + {\omega _{\rm{m}}}{G^2}\Delta {\left( {\frac{{{\gamma _{\rm{m}}}}}{2} + 2\kappa } \right)^2} > 0.\label{eq-29}
\end{eqnarray}
\end{widetext}
The following numerical simulation shows that realistic experimental parameter configurations always meet these stability conditions. When the system is stable, it reaches a unique steady state in the long-time limit $t \to  + \infty $ independently of the initial condition. 

We set the initial to a Gaussian state, and the linear dynamics preserve the noise operators ${\xi '}$, $\xi $, and ${a_{{\rm{in}}}}$. Thus, the correlation properties of the system can be completely characterized by its two first moments, of which we are interested in the second one, namely the covariance matrix with elements defined as
\begin{eqnarray}
{{{V}}_{{{ij}}}} \!&=&\! \frac{1}{2}\left\langle {{\mu _{{i}}}\left( { + \infty } \right){\mu _{{j}}}\left( { + \infty } \right) + {\mu _{{j}}}\left( { + \infty } \right){\mu _{{i}}}\left( { + \infty } \right)} \right\rangle \label{eq-30}\\
 &=& \sum\limits_{{{k}},{{l}}} {\int_{{t_0}}^{ + \infty } \!{d\tau \int_{{t_0}}^{ + \infty } \!{d\tau '{{{M}}_{{{ik}}}}\left( \tau  \right)} } } {{{M}}_{{{jl}}}}\left( {\tau '} \right){\Phi _{{{kl}}}}\left( {\tau  - \tau '} \right),\nonumber
\end{eqnarray}
where ${\Phi _{{{kl}}}}\left( {\tau  - \tau '} \right) = {{\left\langle {{n_{{k}}}\left( \tau  \right){n_{{l}}}\left( {\tau '} \right) + {n_{{l}}}\left( {\tau '} \right){n_{{k}}}\left( \tau  \right)} \right\rangle } \mathord{\left/{\vphantom {{\left\langle {{n_{{k}}}\left( \tau  \right){n_{{l}}}\left( {\tau '} \right) + {n_{{l}}}\left( {\tau '} \right){n_{{k}}}\left( \tau  \right)} \right\rangle } 2}} \right.\kern-\nulldelimiterspace} 2}$ is the matrix of the stationary noise correlation functions. Because the matrix elements are independent of $n\left( t \right)$, we obtain ${\Phi _{{{kl}}}}\left( {\tau  - \tau '} \right) = {{{D}}_{{{kl}}}}\delta \left( {\tau  - \tau '} \right)$, where ${{D}} = {\rm{Diag}}\left[ {{{{\gamma _{\rm{m}}}\left( {2{{\bar n}} + 1} \right)} \mathord{\left/{\vphantom {{{\gamma _{\rm{m}}}\left( {2{{\bar n}} + 1} \right)} 4}} \right.\kern-\nulldelimiterspace} 4},{{{\gamma _{\rm{m}}}\left( {2{{\bar n}} + 1} \right)} \mathord{\left/{\vphantom {{{\gamma _{\rm{m}}}\left( {2{{\bar n}} + 1} \right)} 4}} \right.\kern-\nulldelimiterspace} 4},\kappa ,\kappa } \right]$ is a diagonal matrix. According to Eq.~(\ref{eq-30}) and the form of ${\Phi _{{{kl}}}}\left( {\tau  - \tau '} \right)$, we find that the expression of the matrix ${{V}}$ is equivalent to
\begin{eqnarray}
{{V}} = \int_{{t_0}}^{ + \infty } {{{M}}\left( \tau  \right){{DM}}{{\left( \tau  \right)}^T}d\tau }. \label{eq-31}
\end{eqnarray}
Hence, we obtain
\begin{eqnarray}
  A V & = & \int_{t_0}^{\infty} A M (\tau) D M (\tau)^T d \tau \label{eq-32}\\ &=&
  \int_{t_0}^{\infty} \frac{d}{d \tau} M (\tau) D M (\tau)^T d \tau,\nonumber \\
  V A^T & = & \int_{t_0}^{\infty} M (\tau) D (A M (\tau))^T d \tau \label{eq-33}\\&=&
  \int_{t_0}^{\infty} M (\tau) D \frac{d}{d \tau} M (\tau)^T d \tau.\nonumber
\end{eqnarray}
The combination of Eqs.~(\ref{eq-32}) and (\ref{eq-33}) becomes
\begin{eqnarray}
AV + V{A^T} =  && \!\!\!\!+ \int_{{t_0}}^\infty  {\frac{d}{{d\tau }}} [M(\tau )DM{(\tau )^T}]d\tau \nonumber\\
 && \!\!\!\!- \int_{{t_0}}^\infty  M (\tau )\frac{d}{{d\tau }}DM{(\tau )^T}d\tau \label{eq-34}\\
 = && \!\!\!\! [M\left( \tau \right)DM{{\left( \tau \right)}^T}]|_{{t_0}}^{+\infty} = -D \nonumber
\end{eqnarray}
where we use the assumptions that the stability conditions are satisfied. The solution ${{M}}\left( { + \infty } \right)$ converges to zero in the long-time limit. Equation~(\ref{eq-34}) is a linear Lyapunov equation with respect to ${V}$, which can be solved straightforwardly. See Appendix~\ref{appendix-D} for a detailed derivation of a Lyapunov equation~(\ref{eq-34}). 

We can derive a Lyapunov equation satisfied by the high-frequency inverse-resonance Hamiltonian $H_{{\rm{F}}}^{{\rm{I}}}$ in Eq.~(\ref{eq-12}) similarly to the form of the high-frequency resonance Hamiltonian $H_{{\rm{F}}}$ in Eq.~(\ref{eq-11}). Moreover, we show that the analysis and results concerning optomechanical entanglement in the high-frequency inverse-resonance regime are equivalent to those in the high-frequency resonance regime. Therefore, we do not elaborate on it further here.

\subsection{Optomechanical entanglement}
Cavity optomechanical systems naturally exhibit complex entanglement structures and always involve mixed states and continuous variable entanglement, which are affected by dissipation and noise. In this sense, the logarithmic negativity is a powerful tool that can provide valuable insights into the nature of optomechanical entanglement \cite{ref-59}, which can be experimentally measured using homodyne detection. Thus, we use the logarithmic negativity ${E_{\rm{N}}}$ to measure optomechanical entanglement between the optical cavity field and the mechanical oscillator. It provides an obvious easy way to compute an upper bound for the distillable optomechanical entanglement \cite{ref-60}. 

As mentioned in the continuous variable scenario, the bipartite optomechanical entanglement can be quantified as \cite{ref-56}
\begin{eqnarray}
{E_{\rm{N}}} = \max \left[ {0, - \ln \left( {2{\Xi  }} \right)} \right], \label{eq-35}
\end{eqnarray}
where
\begin{eqnarray}
\Xi  = \frac{1}{{\sqrt 2 }}{\left\{ {\Sigma \left( {{V}} \right) - \sqrt {{{\left[ {\Sigma \left( {{V}} \right)} \right]}^2} - 4\det \left( {{V}} \right)} } \right\}^{\frac{1}{2}}} \label{eq-36}
\end{eqnarray}
\begin{figure}[t]
\centering
\includegraphics[angle=0,width=0.46\textwidth]{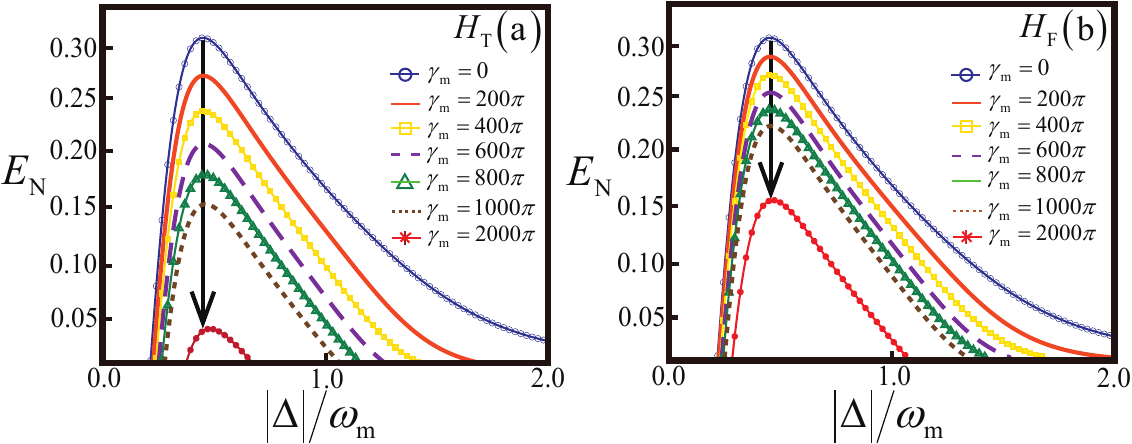}
\caption{ Plot of the logarithmic negativity ${E_{\rm{N}}}$ as a function of the normalized detuning frequency of the optical field $\left| \Delta  \right|$ (in units of ${{\omega _{\rm{m}}}}$) for seven values of the mechanical damping rate: ${\gamma _{\rm{m}}} = 0$ (blue circular line), ${\gamma _{\rm{m}}} = 200\pi {\rm{Hz}}$ (orange solid line), ${\gamma _{\rm{m}}} = 400\pi {\rm{Hz}}$ (yellow square line), ${\gamma _{\rm{m}}} = 600\pi {\rm{Hz}}$ (purple dashed line), ${\gamma _{\rm{m}}} = 800\pi {\rm{Hz}}$ (green triangle line), ${\gamma _{\rm{m}}} = 1000\pi {\rm{Hz}}$ (brown dotted line), and ${\gamma _{\rm{m}}} = 2000\pi {\rm{Hz}}$ (red cross line), where $\left( {\rm{a}} \right)$ and $\left( {\rm{b}} \right)$ correspond to the original model ${H_{\rm{T}}}$ in Eq.~(\ref{eq-10}) and the filtering model ${H_{{\rm{F}}}}$ in Eq.~(\ref{eq-11}), respectively. The length of the black downward-pointing arrows indicates how sensitive optomechanical entanglement is to ${\gamma _{\rm{m}}}$. The other parameters for $\left( {\rm{a}} \right)$ and $\left( {\rm{b}} \right)$ are chosen as follows: the optical cavity of length $L = 1{\rm{mm}}$ and the drives laser with wavelength $\lambda  = 810{\rm{nm}}$ and power $P = 50{\rm{mW}}$. The decay rate of the optical cavity is chosen to be  $\kappa  = 8.8\pi  \times {10^6}{\rm{Hz}}$, the optical finesse $F = \pi c/L\kappa  \approx 3.4 \times {10^4}$ with $c = 3 \times {10^8}{\rm{m}}/{\rm{s}}$, and the driving laser frequency is resonant with the characteristic frequency of the cavity field, ${\omega _L} = {\omega _{\rm{c}}} = 2\pi c/\lambda $. The mechanical oscillator has the characteristic frequency ${\omega _{\rm{m}}} = 20\pi {\rm{MHz}}$, the effective mass $m = 50{\rm{ng}}$, and its temperature is $T = 400{\rm{mK}}$~\cite{ref-24}. \label{Fig-2}}
\end{figure}
is the lowest symplectic eigenvalue of the partial transpose of the $4 \times 4$ steady-state correlation matrix~\cite{ref-61}. For simplicity, we denote the $4 \times 4$ steady-state correlation matrix as in $2 \times 2$ block matrix form, which is given by ${{V}} = \left[ {\left( {\Theta ,\beta } \right),\left( {{\beta ^T},\eta } \right)} \right]$, and $\Sigma \left( {{V}} \right) = \det \left( \Theta  \right) + \det \left( \eta  \right) - 2\det \left( \beta  \right)$. We note that a Gaussian state is entangled if and only if $\Xi  < {1 \mathord{\left/{\vphantom {1 2}} \right.\kern-\nulldelimiterspace} 2}$. It is equivalent to Simon's entanglement criteria for all bipartite Gaussian states \cite{ref-62}, which can be written as $4\det \left( {{V}} \right) < \Sigma {\left( {{V}} \right)}  - {1 \mathord{\left/{\vphantom {1 4}} \right.\kern-\nulldelimiterspace} 4}$.

\begin{figure}[t]
\centering
\includegraphics[angle=0,width=0.46\textwidth]{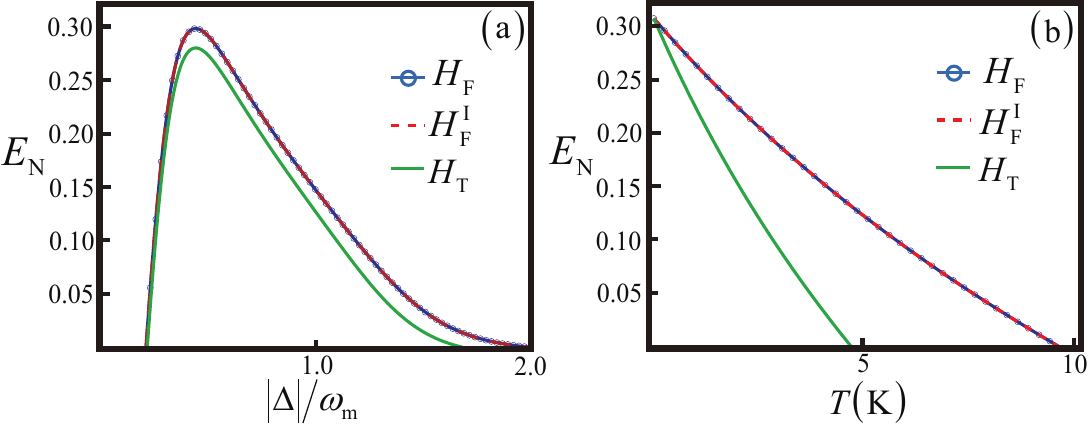}
\caption{ Comparing the optomechanical entanglement properties under different mechanisms, the high-frequency resonance of the filtering model $H_{{\rm{F}}}$ in Eq.~(\ref{eq-11}) (blue circular line), the high-frequency inverse-resonance of the filtering model $H_{{\rm{F}}}^{{\rm{I}}}$ in Eq.~(\ref{eq-12}) (red dashed line), and the original model ${H_{\rm{T}}}$ in Eq.~(\ref{eq-10}) (green solid line). $\left( {\rm{a}} \right)$ Plot of the logarithmic negativity ${E_{\rm{N}}}$ as a function of the normalized detuning frequency of the optical field $\left| \Delta  \right|$ (in units of ${{\omega _{\rm{m}}}}$). We set ${\gamma _{\rm{m}}} = 200\pi {\rm{Hz}}$ and $T = 400{\rm{mK}}$. $\left( {\rm{b}} \right)$ Plot of the logarithmic negativity ${E_{\rm{N}}}$ versus the mirror temperature $T$. We set ${\gamma _{\rm{m}}} = 200\pi {\rm{Hz}}$ and $\Delta  = 0.5{\omega _{\rm{m}}} = 10\pi {\rm{MHz}}$. Both in $\left( {\rm{a}} \right)$ and $\left( {\rm{b}} \right)$, the other parameter values are the same as in Fig.~\ref{Fig-2}.  \label{Fig-3}}
\end{figure}

We numerically calculated the negativity for cavity optomechanical systems as shown in Figs.~\ref{Fig-2} and~\ref{Fig-3}. In our numerical simulation, we utilize the parameter values identical to those outlined in Ref.~\cite{ref-24}, which agree with the current optomechanical experiments configurations \cite{ref-63,ref-64,ref-65,ref-66} and satisfy the stability conditions~(\ref{eq-29}). To begin with, we set the initial closed-optomechanical system in a maximum optomechanical entangled state. For simplicity, we assume that the driving laser frequency ${\omega _L}$ is resonant with the characteristic frequency ${\omega _{\rm{c}}}$ of the cavity field, that is, the laser detuning from the cavity resonance satisfies ${\Delta _0} = 0$. 

In Fig.~\ref{Fig-2}, we compare the sensitivity of the optomechanical entanglement ${E_{\rm{N}}}$ to the mechanical damping rate ${\gamma _{\rm{m}}}$ for the two optomechanical systems, ${H_{\rm{T}}}$ in Eq.~(\ref{eq-10}) and $H_{\rm{F}}$ in Eq.~(\ref{eq-11}). We show a significant enhancement of the robustness of optomechanical entanglement for $H_{{\rm{F}}}$ against ${\gamma _{\rm{m}}}$. Specifically, we observe that the length of the black downward-pointing arrow in Fig.~\ref{Fig-2}$\left( {\rm{b}} \right)$ is approximately half of that in Fig.~\ref{Fig-2}$\left( {\rm{a}} \right)$, which implies that the optomechanical entanglement of the filtering model ${H_{\rm{F}}}$~(\ref{eq-11}) is almost twice as robust to ${\gamma _{\rm{m}}}$ as the original model ${H_{\rm{T}}}$~(\ref{eq-10}). Additionally, it is worth noting that the presence of optomechanical entanglement is only within a limited range of $\left| \Delta  \right|$ around $\left| \Delta  \right| \approx {\omega _{\rm{m}}}$, which means that the frequency resonance between the normalization of the detuning frequency of the optical field $\left| \Delta  \right|$ and the frequency of the mechanical oscillator ${\omega _{\rm{m}}}$ plays a dominant role in the generation of optomechanical entanglement.

We further examine the impact of the resonance effect between the mechanical mode and its thermal reservoir on the properties of optomechanical entanglement. For this purpose, we set ${\gamma _{\rm{m}}} = 200\pi {\rm{Hz}}$ according to the actual laboratory conditions. 

Figure~\ref{Fig-3}$\left( {\rm{a}} \right)$ shows the logarithmic negativity ${E_{\rm{N}}}$ versus the normalized detuning frequency of the optical field $\left| \Delta  \right|$ (in units of ${{\omega _{\rm{m}}}}$) for cases models, the high-frequency resonance of the filtering model $H_{{\rm{F}}}$ in Eq.~(\ref{eq-11}), the high-frequency inverse-resonance of the filtering model $H_{{\rm{F}}}^{{\rm{I}}}$ in Eq.~(\ref{eq-12}), and the original system ${H_{\rm{T}}}$ in Eq.~(\ref{eq-10}). It shows that the maximum optomechanical entanglements for $H_{{\rm{F}}}$ and $H_{{\rm{F}}}^{{\rm{I}}}$ are equal to each other while that for ${H_{\rm{T}}}$ is less than it. The results indicate that the resonance effect can safeguard the maximum optomechanical entanglement by filtering out the contributions from a largely detuned part of the degree of freedom, ultimately reducing both the Brownian noise $\xi $ $\left( {\xi '} \right)$ and the mechanical dissipation ${\gamma _{\rm{m}}}$. 

The robustness of such an entanglement ${E_{\rm{N}}}$ with respect to the environmental temperature $T$ of the mirror is shown in Fig.~\ref{Fig-3}$\left( {\rm{b}} \right)$. We find that the optomechanical entanglement of the filtering model ${H_{\rm{F}}}$ in Eq.~(\ref{eq-11}) remains even at temperatures around 10K and is twice the magnitude of the persistent temperature in the original model~${H_{\rm{T}}}$ in Eq.~(\ref{eq-10}). In addition, we observe that the high-frequency resonance and the high-frequency inverse-resonance regimes have completely equivalent effects on optomechanical entanglement.

In summary, we have discussed the impact of the high-frequency resonance effect between the mechanical oscillator and its thermal reservoir on optomechanical entanglement. We have found that the resonance effect doubles the robustness of optomechanical entanglement to the mechanical dissipation and the mirror temperature. We have achieved the maximum protection of optomechanical entanglement by constructing a filtering model using resonance effects. We have observed numerically that both the high-frequency resonance and the high-frequency inverse-resonance regimes have equivalent effects on optomechanical entanglement.

\subsection{Experimental Implementation}\label{section4}
\begin{figure}[h]
\centering
\includegraphics[angle=0,width=0.48\textwidth]{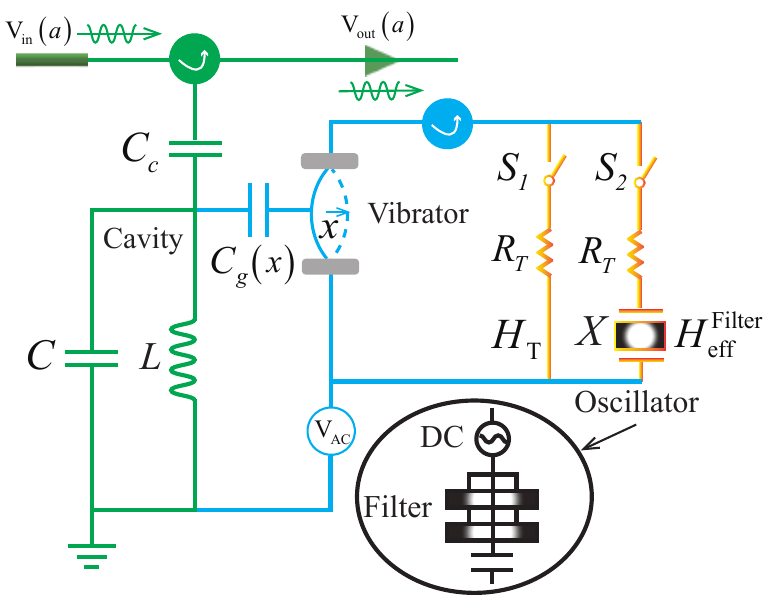}
\caption{ A circuit consisting of a resistor, inductor, and capacitor can be used to build an oscillatory filtering model for high-frequency resonance. This experimental setup comprises an on-chip optical cavity (green) coupled with a high-quality-factor nano-mechanical resonator. By turning on switch 1 and turning off switch 2, the thermistor (orange) will provide a thermal environment that couples with the resonator, corresponding to the original model ${H_{\rm{T}}}$ in Eq.~(\ref{eq-10}). Conversely, the thermistor and the oscillator (black) will generate a high-frequency oscillation thermal environment that couples with the resonator, corresponding to the filtering model $H_{{\rm{F}}}$ in Eq.~(\ref{eq-11}) Direct current, abbreviated as DC, is used for signal frequency readout. \label{Fig-4}}
\end{figure}

\begin{figure*}[t]
\centering
\includegraphics[angle=0,width=0.92\textwidth]{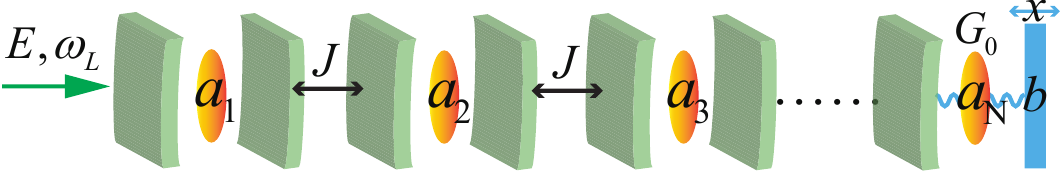}
\caption{ The schematic diagram depicts a one-dimensional array of optical cavities coupled via linear hopping between each cavity, with an oscillating end-mirror.\label{Fig-5}}
\end{figure*}

We propose materializing the present theoretical filtering model in a resistor-inductor-capacitor circuit \cite{ref-67,ref-68,ref-69} or superconducting quantum interference device experiments \cite{ref-70}. As shown in Fig.~\ref{Fig-4}, we build an oscillatory circuit consisting of a capacitor $C$, an inductor $L$, a thermistor ${R_T}$, and an oscillator $X$. We set the normalized detuning frequency of the optical field of the LC circuit to satisfy $\left| \Delta  \right|  = {1 \mathord{/{\vphantom {1 {( {2\pi \sqrt {LC} } )}}} \kern-\nulldelimiterspace} {( {2\pi \sqrt {LC} } )}} = 20\pi {\rm{MHz}}$. First, the mechanical resonator (blue) and the optical cavity (green) are connected via an inductor. Second, an extensive AC voltage bias ${{\rm{V}}_{{\rm{AC}}}}$ is applied in order to excite the mechanical resonator, represented as a movable capacitance ${C_g}\left( x \right)$. Here, to obtain the maximum optomechanical entanglement, the frequency of the applied voltage should be close to $\left| \Delta  \right|$, namely $\left| \Delta  \right| \approx {\omega _{\rm{m}}}$. Next, as the LC circuit oscillates, a current is induced in the thermistor, generating a temperature change due to the Joule heating effect. Therefore, by turning on switch 1 and turning off switch 2 simultaneously, the mechanical resonator will be coupled to a full-frequency thermal reservoir, corresponding to the original model ${H_{\rm{T}}}$ in Eq.~(\ref{eq-10}). In contrast, the largely detuned part of the degree of freedom can be filtered by applying the oscillator $X$ if we turn off switch 1 while turning on switch 2. The oscillator X is an electronic circuit component capable of generating a specific frequency signal and can be utilized as a filter to filter out unwanted frequency components selectively. Specifically, when the input signal matches the resonant frequency of the oscillator, it amplifies the input signal and outputs a near-resonant signal, thereby achieving high-frequency oscillatory wave filtering. Thus, the resistor-inductor-capacitor oscillatory circuit can be described by the filtering model $H_{{\rm{F}}}$ in Eq.~(\ref{eq-11}). 

In addition, we need to choose a mechanical resonator with a giant mechanical quality factor to ensure that significant quantum effects are achievable, that is, $Q = {{{\omega _{\rm{m}}}} \mathord{\left/{\vphantom {{{\omega _{\rm{m}}}} {{\gamma _{\rm{m}}} = {1 \mathord{/{\vphantom {1 \gamma }} \kern-\nulldelimiterspace} \gamma }}}} \right.\kern-\nulldelimiterspace} {{\gamma _{\rm{m}}} = {1 \mathord{/{\vphantom {1 \gamma }} \kern-\nulldelimiterspace} \gamma }}} \gg 1$ corresponding to the weak-coupling limit $\gamma  \ll 1$. The remaining parameter values for the simulation of the circuit experiment are the same as in Fig.~\ref{Fig-3}$\left( {\rm{a}} \right)$. Furthermore, we note that with optical interferometry techniques \cite{ref-71,ref-72}, we can observe the resonance response of a mechanical resonator to its thermal environment. The homodyne detection techniques \cite{ref-73,ref-74} can be used to measure an optomechanical entanglement.

It is important to note that experimental studies on open-system dynamics with linear optical setups often use approximated simulations of quantum channels, such as amplitude decay or phase-damping channels~\cite{ref-75,ref-76,ref-77,ref-78} These simulations rely on the rotating-wave approximation for system-bath interactions and the weak coupling approximation. Recently, we noted that a study aims to test the difference between rotating-wave approximation and non-rotating-wave approximation channels by studying the varying dynamics of quantum temporal steering was demonstrated experimentally~\cite{ref-79,ref-80}.

\section{Generalized Extension and Application}\label{section4}

We are now extending the theory of resonance-dominant entanglement to a multi-mode optomechanical system. Specifically, we discuss an optical-cavity array with one oscillating end mirror and investigate optimal optomechanical entanglement transmission. 

As schematically shown in Fig.~\ref{Fig-5}, the system comprises an oscillating end mirror coupled to an array of optical cavities. The adjacent optical cavities are linearly coupled with an interaction strength of $J$ \cite{ref-81}. A laser field drives the left end of the optical cavity, while the right end is connected to a vibrating end mirror.

If we consider this system satisfying the resonance regime, the total Hamiltonian of this open quantum system can be written as

\begin{eqnarray}
H = \!\!\!\!\!&&+ \hbar {\Delta _0}a_1^\dag {a_1} + \sum\limits_{j = 2}^N {{\omega _{{{\rm{c}}_j}}}} a_j^\dag {a_j} + \hbar {\omega _{\rm{m}}}{b^\dag }b \nonumber\\
 &&+ i\hbar \left( {Ea_1^\dag  - {E^*}{a_1}} \right) + \hbar \sum\limits_{j = 1}^{N - 1} {J\left( {a_j^\dag {a_{j + 1}} + a_{j + 1}^\dag {a_j}} \right)} \nonumber\\
 &&- \hbar \frac{{{G_0}}}{{\sqrt 2 }}a_N^\dag {a_N}\left( {{b^\dag } + b} \right) + \hbar \sum\limits_{j = 1}^N {\sum\limits_k {{\omega _{jk}}\Gamma _{jk}^\dag } } {\Gamma _{jk}} \nonumber\\
 &&+ \hbar \sum\limits_{j = 1}^N {\sum\limits_k {{g_{jk}}\left( {\Gamma _{jk}^\dag {a_j} + a_j^\dag {\Gamma _{jk}}} \right)} } \label{eq-37}\\
 &&+ \hbar \sum\limits_n {{\omega _n}\Lambda _n^\dag } {\Lambda _n} - i\hbar \sum\limits_n {\frac{{{\sigma _n}}}{2}} \left( {\Lambda _n^\dag b - {b^\dag }{\Lambda _n}} \right),\nonumber
\end{eqnarray}
where ${a_{{j}}^\dag }$ (${{a_{{j}}}}$) and ${\Gamma _{{{jk}}}^\dag }$ (${{\Gamma _{{{jk}}}}}$) are the corresponding creation (annihilation) operators for the ${{j}}$th optical cavity mode and its thermal reservoir modes with frequencies ${{\omega _{{{\rm{c}}_{{j}}}}}}$ and ${{\omega _{{{jk}}}}}$, respectively, and the coupling strength between them is ${{g_{{{jk}}}}}$. 

Similarly, nonlinear Langevin equations for the operators of the mechanical and optical modes are given as follows:
\begin{eqnarray}
\dot q \!&=&\! {\omega _{\rm{m}}}p - \frac{{{\gamma _{\rm{m}}}}}{4}q + \frac{1}{2}\xi ',\nonumber\\
\dot p \!&=&\!  - {\omega _{\rm{m}}}q - \frac{{{\gamma _{\rm{m}}}}}{4}p + {G_0}a_{\rm{N}}^\dag {a_{\rm{N}}} + \frac{1}{2}\xi ,\nonumber\\
{{\dot a}_1} \!&=&\!  - \left( {\kappa  + i{\Delta _0}} \right){a_1} - iJ{a_2} + E + \sqrt {2\kappa } a_1^{{\rm{in}}}, \cdots ,\label{eq-38}\\
{{\dot a}_{{j}}} \!&=&\!  - \left( {\kappa  + i{\omega _{{{\rm{c}}_{{j}}}}}} \right){a_{{j}}} - iJ\left( {{a_{{{j}} - 1}} + {a_{{{j + }}1}}} \right) + \sqrt {2\kappa } a_{{j}}^{{\rm{in}}}, \cdots ,\nonumber\\
{{\dot a}_{{N}}} \!&=&\!  - \left( {\kappa  + i{\omega _{{{\rm{c}}_{{N}}}}}} \right){a_{{N}}} - iJ{a_{{{N - 1}}}} + i{G_0}q{a_{{N}}} + \sqrt {2\kappa } a_{{N}}^{{\rm{in}}}, \nonumber
\end{eqnarray}
where we assume that all optical-cavity fields share the same coupling strength: ${g_{{{jk}}}} = {g_{{k}}}$, i.e., ${\kappa _{{j}}} = \kappa $. As the simplest case, we consider ${{N}} = 2$ to study the optomechanical entanglement properties of this system. Similarly, we use the logarithmic negativity to measure the entanglement between two arbitrary bosonic modes in the system. Now, we focus on the numerical evaluation of the bipartite entanglement $E_{\rm{N}}^{{\rm{mc}} \cdot {\rm{1}}}$ to show the optimal remote optomechanical entanglement transfer.

\begin{figure}[htbp]
\centering
\includegraphics[angle=0,width=0.46\textwidth]{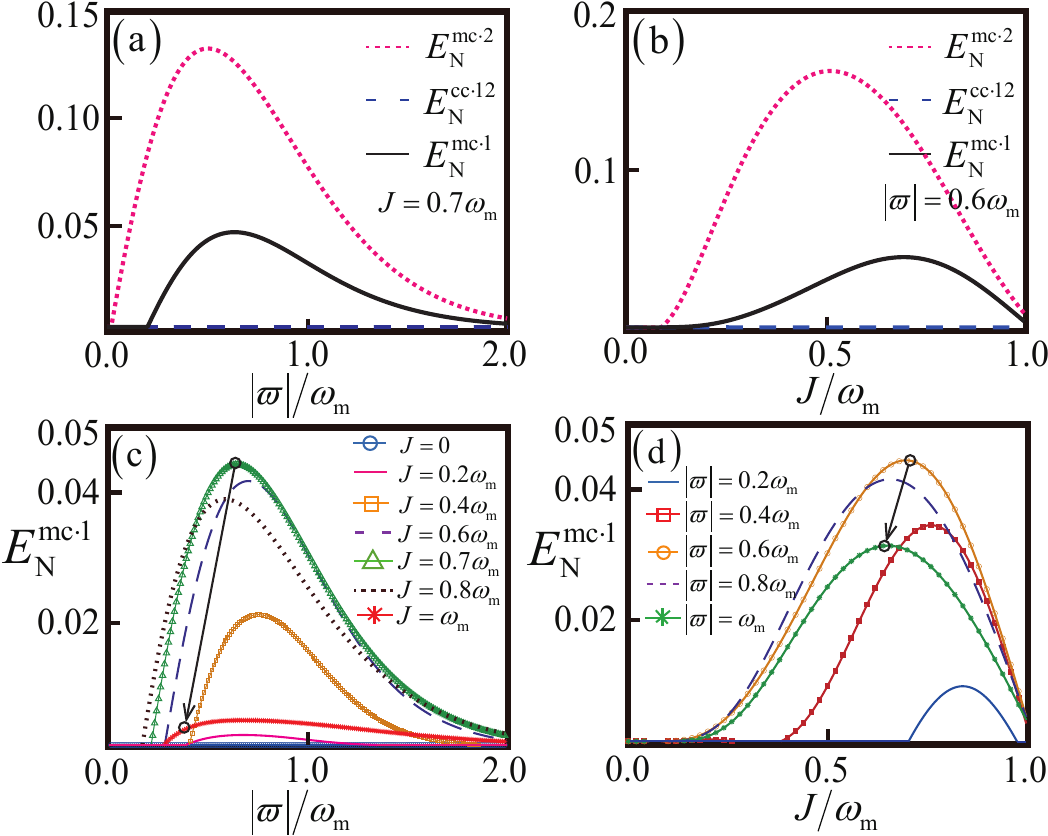}
\caption{The optimal remote optomechanical entanglement transmission. $\left( {\rm{a}} \right)$ The negativity entanglements $E_{\rm{N}}^{{\rm{mc}} \cdot {\rm{1}}}$ (pink dotted line), $E_{\rm{N}}^{{\rm{mc}} \cdot {\rm{2}}}$ (black solid line), and $E_{\rm{N}}^{{\rm{cc}} \cdot {\rm{12}}}$ (blue dashed line) as a function of the normalized detuning $\left| \varpi  \right|$ (in units of ${{\omega _{\rm{m}}}}$) with the other parameters set to ${\Delta _0} = 0$ and $J = 0.7{\omega _{\rm{m}}}$. $\left( {\rm{b}} \right)$ The negativity entanglements $E_{\rm{N}}^{{\rm{mc}} \cdot {\rm{1}}}$, $E_{\rm{N}}^{{\rm{mc}} \cdot {\rm{2}}}$, and $E_{\rm{N}}^{{\rm{cc}} \cdot {\rm{12}}}$ versus the linear hopping strength $J$ (in units of ${{\omega _{\rm{m}}}}$) with the other parameters set to ${\Delta _0} = 0$ and $\left| \varpi  \right|  = 0.6{\omega _{\rm{m}}}$. $\left( {\rm{c}} \right)$ The negativity entanglements $E_{\rm{N}}^{{\rm{mc}} \cdot {\rm{1}}}$ as a function of $\left| \varpi  \right|$ (in units of ${{\omega _{\rm{m}}}}$) for different values of the linear hopping rate: $J = 0$ (blue circular line), $J = 0.2{\omega _{\rm{m}}}$ (pink solid line), $J = 0.4{\omega _{\rm{m}}}$ (orange square line), $J = 0.6{\omega _{\rm{m}}}$ (purple dashed line), $J = 0.7{\omega _{\rm{m}}}$ (green triangle line), $J = 0.8{\omega _{\rm{m}}}$ (brown dotted line), and $J = {\omega _{\rm{m}}}$ (red cross line). $\left( {\rm{d}} \right)$ The negativity entanglements $E_{\rm{N}}^{{\rm{mc}} \cdot {\rm{1}}}$ as a function of $J$ (in units of ${{\omega _{\rm{m}}}}$) for different values of the normalized detuning: $\left| \varpi  \right|  = 0.2{\omega _{\rm{m}}}$ (blue solid line), $\left| \varpi  \right|  = 0.4{\omega _{\rm{m}}}$ (red square line), $\left| \varpi  \right|  = 0.6{\omega _{\rm{m}}}$ (orange circular line), $\left| \varpi  \right|  = 0.8{\omega _{\rm{m}}}$ (purple dashed line), and $\left| \varpi  \right|  = {\omega _{\rm{m}}}$ (green cross line). The remaining parameter values for $\left( {\rm{a}} \right)$-$\left( {\rm{d}} \right)$ are set to be the same as in Fig.~\ref{Fig-3}.  \label{Fig-6}}
\end{figure}

In the two-cavity case, we let $E_{\rm{N}}^{{\rm{mc}} \cdot {\rm{1}}}$, $E_{\rm{N}}^{{\rm{mc}} \cdot {\rm{2}}}$, and $E_{\rm{N}}^{{\rm{cc}} \cdot {\rm{12}}}$ denote the logarithmic negativity between the mirror and the cavity 1, the mirror and the cavity 2, and the cavity 1 and the cavity 2, respectively. In Fig.~\ref{Fig-6}$\left( {\rm{a}} \right)$, we plot $E_{\rm{N}}^{{\rm{mc}} \cdot {\rm{1}}}$, $E_{\rm{N}}^{{\rm{mc}} \cdot {\rm{2}}}$, and $E_{\rm{N}}^{{\rm{cc}} \cdot {\rm{12}}}$ as functions of the normalized detuning $\left| \varpi  \right|$ (in units of ${{\omega _{\rm{m}}}}$) with the other parameters set to ${\Delta _0} = 0$, $J = 0.7{\omega _{\rm{m}}}$, and $T = 400{\rm{mK}}$. The normalized detuning $\varpi  = {\omega _{{{\rm{c}}_2}}} - {G_0}{Q_{\rm{s}}}$ depends on the steady-state mean values ${Q_{\rm{s}}} = {G_0}\alpha _{2{\rm{s}}}^*{\alpha _{2{\rm{s}}}}/{\omega _{\rm{m}}}$, and ${\alpha _{2{\rm{s}}}} =  - iJ{\alpha _{{\rm{1s}}}}/\left( {\kappa  + i\varpi } \right)$ with ${\alpha _{{\rm{1s}}}} = E/\left[ {\kappa  + i{\Delta _0} + {J^2}/\left( {\kappa  + i\varpi } \right)} \right]$, which can be obtained by setting the time derivation to zero in the nonlinear Langevin equation~(\ref{eq-38}) for ${{N}} = 2$. Our numerical findings show that by tuning the magnitude of $\varpi $, we are able to achieve long-distance optomechanical-entanglement transfer. As $\left| \varpi  \right|$ increases approximately from $0.50{\omega _{\rm{m}}}$ to $0.65{\omega _{\rm{m}}}$, the distant optomechanical entanglement $E_{\rm{N}}^{{\rm{mc}} \cdot {\rm{1}}}$ correspondingly increases at the expense of the decrease of the neighboring optomechanical entanglement $E_{\rm{N}}^{{\rm{mc}} \cdot {\rm{2}}}$, due to the adjacent cavities acting as entanglement transmitters. 

In Fig.~\ref{Fig-6}$\left( {\rm{b}} \right)$, we plot $E_{\rm{N}}^{{\rm{mc}} \cdot {\rm{1}}}$, $E_{\rm{N}}^{{\rm{mc}} \cdot {\rm{2}}}$, and $E_{\rm{N}}^{{\rm{cc}} \cdot {\rm{12}}}$ as functions of the linear hopping strength $J$ (in units of ${{\omega _{\rm{m}}}}$) with the other parameters set to ${\Delta _0} = 0$, $\left| \varpi  \right|  = 0.6{\omega _{\rm{m}}}$, and $T = 400{\rm{mK}}$. In a similar analysis, we can also implement distant optomechanical entanglement transfer by adjusting the strength of $J$ approximately from $0.5{\omega _{\rm{m}}}$ to $0.75{\omega _{\rm{m}}}$. In particular, when $T = 400{\rm{mK}}$, we find that the optimal remote optomechanical entanglement transfer occurs around $\left| \varpi  \right|  = 0.6$ and $J = 0.7$ (in units of ${{\omega _{\rm{m}}}}$), and the maximum value of remote entanglement $E_{\rm{N}}^{{\rm{mc}} \cdot {\rm{1}}}$ is approximately evaluated at $0.045$; see Fig.~\ref{Fig-6}$\left( {\rm{c}} \right)$-$\left( {\rm{d}} \right)$.

\section{Summary and Prospect}\label{section5}
In summary, we have demonstrated that resonance effects between a mechanical mode and its thermal environment can protect optomechanical entanglement. Specifically, we have shown that resonance effects nearly double the robustness of the optomechanical entanglement against mechanical dissipation and its environmental temperature. The mechanism of optomechanical-entanglement protection involves the elimination of degrees of freedom associated with significant detuning between the mechanical mode and its thermal reservoirs, thereby counteracting the decoherence. We have revealed that this approach is particularly effective when both near-resonant and weak-coupling conditions are simultaneously satisfied between a mechanical mode and its environment. We have also proposed a feasible experimental implementation for the filtering model to observe these phenomena. Furthermore, we extended this theory to an optical cavity array with one oscillating end mirror and investigated optimal optomechanical entanglement transfer. This study represents a significant advancement in the application of resonance effects for protecting quantum systems against decoherence, thereby opening up new possibilities for large-scale quantum information processing and the construction of quantum networks.

In addition, extending the resonance-dominant entanglement theory to non-Markovian and non-Hermitian optomechanical systems is also challenging and expected to be impactful. Specifically, we ensure that studying non-Markovian effects \cite{ref-82,ref-83,ref-84,ref-85}, exceptional points \cite{ref-86}, parity-time symmetry \cite{ref-87}, and anti-parity time symmetry \cite{ref-88} on optomechanical entanglement is exciting. In particular, we are interested in future investigations of the optomechanical entanglement properties between resonance states \cite{ref-89,ref-90} in non-Hermitian systems. This work aims to develop an innovative approach for protecting continuous variable entanglement.

\section*{Acknowledgments}
We are thankful to Naomichi Hatano for his inspiration and careful reading of the manuscript. Cheng Shang has a pleasure to discuss with Kinkawa Hayato. We thank the feasibility suggestions provided by Chun-Hua Dong, Mai Zhang, Zhen Shen, and Yu Wang in the experimental implementation. Cheng Shang acknowledges the financial support by the China Scholarship Council and the Japanese Government (Monbukagakusho-MEXT) Scholarship under Grant No. 211501. Additionally, Cheng Shang would like to acknowledge the financial support provided by the RIKEN Junior Research Associate Program. The other author, Hongchao Li, is supported by the Forefront Physics and Mathematics Program to Drive Transformation (FoPM) and the World-leading Innovative Graduate Study (WINGS) program at the University of Tokyo.

\begin{widetext}
\appendix
\section{Derivation of the Hamiltonian~(\ref{eq-1})} \label{appendix-A}
Here, we show the origin of the total Hamiltonian~(\ref{eq-1})~\cite{ref-43}. The total Hamiltonian~(\ref{eq-1}) of this field reservoir consists of two parts, the system~(\ref{eq-2}) and the environment~(\ref{eq-3}). Therefore, to obtain Eq.~(\ref{eq-1}), we need to demonstrate the specific origins of Eqs.~(\ref{eq-2}) and (\ref{eq-3}) separately.

To begin with, we show the origin of the system Hamiltonian~(\ref{eq-2}). As usual, for an optomechanical system driven by an optical laser, the Hamiltonian of the composite system can be written as
\begin{eqnarray}
H_{\rm{S}}^0 = \hbar {\omega _{\rm{c}}}{a^\dag }a + \frac{{{{p'}^2}}}{{2m}} + \frac{1}{2}m{\left( {{\omega _{\rm{m}}}q'} \right)^2} - \hbar G{a^\dag }aq' + i\hbar \left( {E{e^{ - i{\omega _0}t}}{a^\dag } - {E^*}{e^{i{\omega _0}t}}a} \right),\label{eq-A1}
\end{eqnarray}
where a monochromatic field drives the optical mode with the driving frequency ${{\omega _0}}$, and the complex amplitude of the driving laser is denoted by $E$. The optical frequency shift per displacement is given by $G =  - {{\partial {\omega _{\rm{c}}}\left( x \right)} \mathord{\left/{\vphantom {{\partial {\omega _{\rm{c}}}\left( x \right)} {\partial x}}} \right.\kern-\nulldelimiterspace} {\partial x}} = {{{\omega _{\rm{c}}}} \mathord{\left/{\vphantom {{{\omega _{\rm{c}}}} L}} \right.\kern-\nulldelimiterspace} L}$. To make the Hamiltonian independent of time, we then move to the rotating frame of the frequency, which makes Eq.~(\ref{eq-A1}) as follows:
\begin{eqnarray}
{H'_{\rm{S}}} &=& U\left( t \right)H_{\rm{S}}^0\left( t \right){U^\dag }\left( t \right) - iU\left( t \right){{\dot U}^\dag }\left( t \right) \nonumber \\
 &=& \hbar {\Delta _0}{a^\dag }a + \frac{{{{p'}^2}}}{{2{{m}}}} + \frac{1}{2}{{m}}{\left( {{\omega _{\rm{m}}}q'} \right)^2} - \hbar {G}{a^\dag }aq' + i\hbar \left( {E{a^\dag } - {E^*}a} \right),\label{eq-A2}
\end{eqnarray}
where we used the unitary transformation of the form $U\left( t \right) = \exp \left( {i{\omega _0}{a^\dag }at} \right)$, and ${\Delta _0} = {\omega _{\rm{c}}} - {\omega _0}$ is the detuning of the cavity characteristic frequency ${\omega _{\rm{c}}}$ of the optical cavity from the driving laser frequency ${\omega _0}$.

We can make the position and momentum operators dimensionless by defining the zero-point fluctuation amplitude of the mechanical oscillator as ${{{X}}_{{\rm{ZPF}}}} = \sqrt {{\hbar  \mathord{\left/{\vphantom {\hbar  {2m{\omega _{\rm{m}}}}}} \right.\kern-\nulldelimiterspace} {2m{\omega _{\rm{m}}}}}} $. Then, we define the dimensionless position operator $q$ and momentum operator $p$ as follows:
\begin{eqnarray}
q = \frac{{q'}}{{\sqrt 2 {{ X}_{{\rm{ZPF}}}}}} = \frac{1}{{\sqrt 2 }}\left( {{b^\dag } + b } \right), \quad p = \frac{{p'}}{{\sqrt 2 {{m}}{\omega _{\rm{m}}}{{X}_{{\rm{ZPF}}}}}} = \frac{i}{{\sqrt 2 }}\left( {{b^\dag } - b} \right).\label{eq-A3}
\end{eqnarray}
Substituting Eq.~(\ref{eq-A3}) into Eq.~(\ref{eq-A2}), we arrive at
\begin{eqnarray}
{H_{\rm{S}}} &=& \hbar {\Delta _0}{a^\dag }a + \frac{\hbar }{2}{\omega _{\rm{m}}}\left( {{p^2} + {q^2}} \right) - \hbar {G_0}{a^\dag }aq + i\hbar \left( {E{a^\dag } - {E^*}a} \right) \nonumber \\
 &=& \hbar {\Delta _0}{a^\dag }a + \hbar {\omega _{\rm{m}}}{b^\dag }b - \hbar {G_0}{a^\dag }a\frac{{\left( {{b^\dag } + b} \right)}}{{\sqrt 2 }} + i\hbar \left( {E{a^\dag } - {E^*}a} \right),\label{eq-A4}
\end{eqnarray}
where ${G_0} = \sqrt 2 G{{{X}}_{{\rm{ZPF}}}} = {{{\omega _{\rm{c}}}\sqrt {{\hbar  \mathord{\left/{\vphantom {\hbar  {m{\omega _{\rm{m}}}}}} \right.\kern-\nulldelimiterspace} {m{\omega _{\rm{m}}}}}} } \mathord{/{\vphantom {{{\omega _{\rm{c}}}\sqrt {{\hbar  \mathord{\left/{\vphantom {\hbar  {m{\omega _{\rm{m}}}}}} \right.\kern-\nulldelimiterspace} {m{\omega _{\rm{m}}}}}} } L}} \kern-\nulldelimiterspace} L}$ is the vacuum optomechanical coupling strength, expressed as a frequency. It quantifies the interaction between a single phonon and a single photon. This produces Eq.~(\ref{eq-2}) in the main text.

Next, we give the origin of the environment Hamiltonian~(\ref{eq-3}) for the first time. As is well known from the Bose-Einstein statistics, a heat bath associated with a boson system can be considered as an assembly of harmonic oscillators. This type of heat bath can serve as a model for various physical systems, such as elastic solids (mechanical reservoirs) and electromagnetic fields (optical reservoirs). 

Firstly, since in the optomechanical system, both the photons in the optical cavity and the phonons in the mechanical oscillator obey the Bose-Einstein statistics, the free part of the environment can be written in the simple form
\begin{eqnarray}
H_{\rm{E}}^0 = \frac{1}{2}\sum\limits_{{k}} {\left[ {\frac{1}{{m_{{k}}^{\rm{c}}}}{{\left( {\tilde p_{{k}}^{\rm{c}}} \right)}^2} + \Theta _{{k}}^{\rm{c}}{{\left( {\tilde q_{{k}}^{\rm{c}}} \right)}^2}} \right]}  + \frac{1}{2}\sum\limits_{{n}} {\left[ {\frac{1}{{m_{{n}}^{\rm{m}}}}{{\left( {\tilde p_{{n}}^{\rm{m}}} \right)}^2} + \Theta _{{n}}^{\rm{m}}{{\left( {\tilde q_{{n}}^{\rm{m}}} \right)}^2}} \right]}
,\label{eq-A5}
\end{eqnarray}
where ${m_{{k}}^{\rm{c}}}$ and ${m_{{n}}^{\rm{m}}}$ correspond to the effective mass of the $k$th optical reservoir and $n$th mechanical reservoir, respectively. The momentum and position operators corresponding to the $k$th optical reservoir and the $n$th mechanical reservoir are denoted by ${\tilde p_{{k}}^{\rm{c}}}$ ${\tilde p_{{n}}^{\rm{m}}}$ and ${\tilde q_{{k}}^{\rm{c}}}$ ${\tilde q_{{n}}^{\rm{m}}}$, respectively. We set $\Theta _{{k}}^{\rm{c}} = m_{{k}}^{\rm{c}}{\left( {{\omega _{{k}}}} \right)^2}$ and $\Theta _{{n}}^{\rm{m}} = m_{{n}}^{\rm{m}}{\left( {{\omega _{{n}}}} \right)^2}$ as the optical and mechanical potential-force constants. The harmonic-oscillator reservoirs have closely spaced frequencies corresponding to photons and phonons, denoted by ${\omega _{k}}$ and ${{\omega _{n}}}$, respectively. Through the process of removing the dimensions from the operators, we can define the dimensionless momentum operators $p_{{k}}^{\rm{c}}$ and ${ {p_{{n}}^{\rm{m}}}}$ as well as position operators ${q_{{k}}^{\rm{c}}}$ and ${{q_{{n}}^{\rm{m}}}}$ as follows:
\begin{eqnarray}
p_{{k}}^{\rm{c}} &=& \sqrt {\frac{{{\omega _{{k}}}}}{{\Theta _{{k}}^{\rm{c}}\hbar }}} \tilde p_{{k}}^{\rm{c}} = \sqrt {\frac{1}{{m_{{k}}^{\rm{c}}{\omega _{{k}}}\hbar }}} \tilde p_{{k}}^{\rm{c}}, \qquad \ q_{{k}}^{\rm{c}} = \sqrt {\frac{{\Theta _{{k}}^{\rm{c}}}}{{{\omega _{{k}}}\hbar }}} \tilde q_{{k}}^{\rm{c}} = \sqrt {\frac{{{\omega _{{k}}}}}{\hbar }} \tilde p_{{k}}^{\rm{c}},\label{eq-A6}\\
p_{{n}}^{\rm{m}} &=& \sqrt {\frac{{{\omega _{{n}}}}}{{\Theta _{{n}}^{\rm{m}}\hbar }}} \tilde p_{{n}}^{\rm{m}} = \sqrt {\frac{1}{{m_{{n}}^{\rm{m}}{\omega _{{n}}}\hbar }}} \tilde p_{{n}}^{\rm{m}}, \quad q_{{n}}^{\rm{m}} = \sqrt {\frac{{\Theta _{{n}}^{\rm{m}}}}{{{\omega _{{n}}}\hbar }}} \tilde q_{{n}}^{\rm{m}} = \sqrt {\frac{{{\omega _{{n}}}}}{\hbar }} \tilde p_{{n}}^{\rm{m}}.\label{eq-A7}
\end{eqnarray}
Substituting Eqs.~(\ref{eq-A6}) and~(\ref{eq-A7}) into Eq.~(\ref{eq-A5}), we have
\begin{eqnarray}
{H'_{\rm{E}}} = \frac{\hbar }{2}\sum\limits_{{k}} {{\omega _{{k}}}\left[ {{{\left( {p_{{k}}^{\rm{c}}} \right)}^2} + {{\left( {q_{{k}}^{\rm{c}}} \right)}^2}} \right]}  + \frac{\hbar }{2}\sum\limits_{{n}} {{\omega _{{n}}}\left[ {{{\left( {p_{{n}}^{\rm{m}}} \right)}^2} + {{\left( {q_{{n}}^{\rm{m}}} \right)}^2}} \right]},\label{eq-A8}
\end{eqnarray}

Secondly, we consider the coupling between the system and the environment. The  Hamiltonian of a system can be left arbitrary, such as an atom, as in quantum optics, or a macroscopic LC-circuit. In our case, we treat the optomechanical system as a perturbation to the baths, by writing
\begin{eqnarray}
{H''_{\rm{E}}} &=& + \frac{\hbar }{2}\sum\limits_{{k}} {{\omega _{{k}}}\left[ {{{\left( {p_{{k}}^{\rm{c}}} \right)}^2} + {{\left( {q_{{k}}^{\rm{c}} + \varepsilon _{{k}}^{\rm{c}}{q_{\rm{c}}}} \right)}^2}} \right]}  + \frac{\hbar }{2}\sum\limits_{{n}} {{\omega _{{n}}}\left[ {{{\left( {p_{{n}}^{\rm{m}} - \chi _{{n}}^{\rm{m}}{q_{\rm{m}}}} \right)}^2} + {{\left( {q_{{n}}^{\rm{m}}} \right)}^2}} \right]} \nonumber\\
 &=& + \frac{\hbar }{2}\sum\limits_{{k}} {{\omega _{{k}}}\left[ {{{\left( {p_{{k}}^{\rm{c}}} \right)}^2} + {{\left( {q_{{k}}^{\rm{c}}} \right)}^2}} \right]}  + \frac{\hbar }{2}\sum\limits_{{k}} {{\omega _{{k}}}} {\left( {\varepsilon _{{k}}^{\rm{c}}{q_{\rm{c}}}} \right)^2} + \hbar \sum\limits_{{k}} {{\omega _{{k}}}} \varepsilon _{{k}}^{\rm{c}}q_{{k}}^{\rm{c}}{q_{\rm{c}}} \label{eq-A9} \\
 &&+ \frac{\hbar }{2}\sum\limits_{{n}} {{\omega _{{n}}}\left[ {{{\left( {p_{{n}}^{\rm{m}}} \right)}^2} + {{\left( {q_{{n}}^{\rm{m}}} \right)}^2}} \right]}  + \frac{\hbar }{2}\sum\limits_{{n}} {{\omega _{{n}}}{{\left( {\chi _{{n}}^{\rm{m}}{q_{\rm{m}}}} \right)}^2} - } \hbar \sum\limits_{{n}} {{\omega _{{n}}}} \chi _{{n}}^{\rm{m}}p_{{n}}^{\rm{m}}{q_{\rm{m}}},\nonumber
\end{eqnarray}
or
\begin{eqnarray}
{{\tilde H''}_{\rm{E}}} &=& + \frac{\hbar }{2}\sum\limits_{{k}} {{\omega _{{k}}}\left[ {{{\left( {p_{{k}}^{\rm{c}} + \varepsilon _{{k}}^{\rm{c}}{p_{\rm{c}}}} \right)}^2} + {{\left( {q_{{k}}^{\rm{c}}} \right)}^2}} \right]}  + \frac{\hbar }{2}\sum\limits_{\rm{n}} {{\omega _{{n}}}\left[ {{{\left( {p_{{n}}^{\rm{m}} - \chi _{{n}}^{\rm{m}}{q_{\rm{m}}}} \right)}^2} + {{\left( {q_{{n}}^{\rm{m}}} \right)}^2}} \right]} \nonumber\\
 &=& + \frac{\hbar }{2}\sum\limits_{\rm{k}} {{\omega _{{k}}}\left[ {{{\left( {p_{{k}}^{\rm{c}}} \right)}^2} + {{\left( {q_{{k}}^{\rm{c}}} \right)}^2}} \right]}  + \frac{\hbar }{2}\sum\limits_{{k}} {{\omega _{{k}}}} {\left( {\varepsilon _{{k}}^{\rm{c}}{p_{\rm{c}}}} \right)^2} + \hbar \sum\limits_{{k}} {{\omega _{{k}}}} \varepsilon _{{k}}^{\rm{c}}p_{{k}}^{\rm{c}}{p_{\rm{c}}}\label{eq-A10}\\
 &&+ \frac{\hbar }{2}\sum\limits_{{n}} {{\omega _{{n}}}\left[ {{{\left( {p_{{n}}^{\rm{m}}} \right)}^2} + {{\left( {q_{{n}}^{\rm{m}}} \right)}^2}} \right]}  + \frac{\hbar }{2}\sum\limits_{{n}} {{\omega _{{n}}}{{\left( {\chi _{{n}}^{\rm{m}}{q_{\rm{m}}}} \right)}^2} - } \hbar \sum\limits_{{n}} {{\omega _{{n}}}} \chi _{{n}}^{\rm{m}}p_{{n}}^{\rm{m}}{q_{\rm{m}}}.\nonumber
\end{eqnarray}
The orthogonal relationship for the dimensionless position and momentum operators of the system and the environment read
\begin{eqnarray}
{q_{\rm{c}}} &=& \frac{1}{{\sqrt 2 }}\left( {{a^\dag } + a} \right), \
{p_{\rm{c}}} = \frac{i}{{\sqrt 2 }}\left( {{a^\dag } - a} \right); \quad \ q_{{k}}^{\rm{c}} = \frac{1}{{\sqrt 2 }}\left( {\Gamma _{{k}}^\dag  + {\Gamma _{{k}}}} \right), \ p_{{k}}^{\rm{c}} = \frac{i}{{\sqrt 2 }}\left( {\Gamma _{{k}}^\dag  - {\Gamma _{{k}}}} \right)\!; \label{eq-A11}\\
{q_{\rm{m}}} &=& \frac{1}{{\sqrt 2 }}\left( {{b^\dag } + b} \right), \ \ {p_{\rm{c}}} = \frac{i}{{\sqrt 2 }}\left( {{b^\dag } - b} \right); \quad  q_{{n}}^{\rm{m}} = \frac{1}{{\sqrt 2 }}\left( {\Lambda _{{n}}^\dag  + {\Lambda _{{n}}}} \right), \ p_{{n}}^{\rm{m}} = \frac{i}{{\sqrt 2 }}\left( {\Lambda _{{n}}^\dag  - {\Lambda _{{n}}}} \right)\!. \label{eq-A12}
\end{eqnarray}
By substituting Eqs.~(\ref{eq-A11})-(\ref{eq-A12}) into the Eqs.~(\ref{eq-A9})-(\ref{eq-A10}) and absorbing terms only of the system operators $0.5\hbar \sum\nolimits_{\rm{k}} {{\omega _{{k}}}{{\left( {\varepsilon _{{k}}^{\rm{c}}{p_{\rm{c}}}} \right)}^2}} $, $0.5\hbar \sum\nolimits_{{k}} {{\omega _{{k}}}{{\left( {\varepsilon _{{k}}^{\rm{c}}{q_{\rm{c}}}} \right)}^2}} $, and $0.5\hbar \sum\nolimits_{{n}} {{\omega _{{n}}}{{\left( {\chi _{{n}}^{\rm{m}}{q_{\rm{m}}}} \right)}^2}} $ into the system Hamiltonian, and further neglecting these higher-order perturbations quantities containing ${\left( {\varepsilon _{{k}}^{\rm{c}}} \right)^2}$ and ${\left( {\chi _{{n}}^{\rm{m}}} \right)^2}$, we obtain
\begin{eqnarray}
{{\tilde H''}_{\rm{E}}} \mapsto H_{\rm{E}}^{{q_{\rm{c}}}} &=& +\hbar \sum\limits_{{k}} {{\omega _{{k}}}} \Gamma _{{k}}^\dag {\Gamma _{{k}}} + \hbar \sum\limits_{{k}} {{g_{{k}}}} \left( {\Gamma _{{k}}^\dag {a^\dag } + {\Gamma _{{k}}}a} \right) + \hbar \sum\limits_{{k}} {{g_{{k}}}} \left( {\Gamma _{{k}}^\dag a + {\Gamma _{{k}}}{a^\dag }} \right)\nonumber\\
 &&+ \hbar \sum\limits_{{n}} {{\omega _{{n}}}} \Lambda _{{n}}^\dag {\Lambda _{{n}}} - i\hbar \sum\limits_{{n}} {\frac{{{\sigma _{{n}}}}}{2}} \left( {\Lambda _{{n}}^\dag  - {\Lambda _{{n}}}} \right)\left( {{b^\dag } + b} \right),\label{eq-A13}\\
{{\tilde H''}_{\rm{E}}} \mapsto H_{\rm{E}}^{{p_{\rm{c}}}} &=& +\hbar \sum\limits_{{k}} {{\omega _{{k}}}} \Gamma _{{k}}^\dag {\Gamma _{{k}}} - \hbar \sum\limits_{{k}} {{g_{{k}}}} \left( {\Gamma _{{k}}^\dag {a^\dag } + {\Gamma _{{k}}}a} \right) + \hbar \sum\limits_{{k}} {{g_{{k}}}} \left( {\Gamma _{{k}}^\dag a + {\Gamma _{{k}}}{a^\dag }} \right)\nonumber\\
 &&+ \hbar \sum\limits_{{n}} {{\omega _{{n}}}} \Lambda _{{n}}^\dag {\Lambda _{{n}}} - i\hbar \sum\limits_{{n}} {\frac{{{\sigma _{{n}}}}}{2}} \left( {\Lambda _{{n}}^\dag  - {\Lambda _{{n}}}} \right)\left( {{b^\dag } + b} \right),\label{eq-A14}
\end{eqnarray}
where we set ${g_{{k}}} = 0.5\varepsilon _{{k}}^{\rm{c}}{\omega _{{k}}}$ and ${\sigma _{{n}}} = \chi _{{n}}^{\rm{m}}{\omega _{{n}}}$. The real numbers ${g_{{k}}}$ and ${\sigma _{{n}}}$ represent the coupling strengths between the subsystem and the $n$th reservoir mode, respectively. Finally, we apply the rotating-wave approximation and neglect the counter-rotating terms ${\Gamma _{{k}}^\dag {a^\dag }}$ and ${{\Gamma _{{k}}}a}$ in Eqs.~(\ref{eq-A13}) and~(\ref{eq-A14}), yielding $H_{\rm{E}}^{{q_{\rm{c}}}} \approx \tilde H_{\rm{E}}^{{q_{\rm{c}}}} = {H_{\rm{E}}} = \tilde H_{\rm{E}}^{{p_{\rm{c}}}} \approx H_{\rm{E}}^{{p_{\rm{c}}}}$, where $\tilde H_{\rm{E}}^{{q_{\rm{c}}}}$ and $\tilde H_{\rm{E}}^{{q_{\rm{c}}}}$ represent the Hamiltonian after the rotating-wave approximation. This process produces Eq.~(\ref{eq-3}) in the main text. 

In conclusion, we have physically revealed that photon and phonon perturbations interact with the reservoirs differently. The coupling between photons and the bosonic reservoirs results in the potential energy of the bath depending on the deviation of ${q_{\rm{c}}}$ from all the $q_{{k}}^{\rm{c}}$, while the kinetic energy of the bath depends on the derivation of ${p_{\rm{c}}}$ with respect to all $p_{{k}}^{\rm{c}}$ as well. In other words, it is as if each coordinate $q_{{k}}^{\rm{c}}$ or $p_{{k}}^{\rm{c}}$ is harmonically bound to ${q_{\rm{c}}}$ or ${p_{\rm{c}}}$, respectively. In contrast, the coupling between phonons and the bosonic reservoirs makes the potential energy of the bath depending on the deviation of ${q_{\rm{m}}}$ from all the $p_{{n}}^{\rm{m}}$. The kinetic energy of the bath depends on the derivation of ${p_{\rm{m}}}$ with respect to all $q_{{n}}^{\rm{m}}$ as well. In other words, it is as if each coordinate $q_{{n}}^{\rm{m}}$ or $p_{{n}}^{\rm{m}}$ is harmonically bound to ${p_{\rm{m}}}$ or ${q_{\rm{m}}}$, respectively. In addition, we point out that this difference between perturbations of photons and phonons on the bosonic reservoirs also results in the fact that in the rotating-wave approximation, neglecting the rotating-wave terms ${\Gamma _{{k}}^\dag {a^\dag }}$ and ${{\Gamma _{{k}}}a}$ in the coupling between photons and the electromagnetic field leads to the simplification of $\sum\nolimits_{{k}} {{\omega _{{k}}}[ {{{\left( {p_{{k}}^{\rm{c}} + \varepsilon _{{k}}^{\rm{c}}{p_{\rm{c}}}} \right)}^2} + {{\left( {q_{{k}}^{\rm{c}}} \right)}^2}}] \approx } \sum\nolimits_{{k}} {{\omega _{{k}}}[ {{{\left( {p_{{k}}^{\rm{c}}} \right)}^2} + {{\left( {q_{{k}}^{\rm{c}} + \varepsilon _{{k}}^{\rm{c}}{q_{\rm{c}}}} \right)}^2}} ]} $, while neglecting the counter-rotating terms $\Lambda _{{n}}^\dag b$ and ${\Lambda _{{n}}}{b^\dag }$ in the coupling between phonons and elastic solid simplifies $\sum\nolimits_{{n}} {{\omega _{{n}}}[ {{{\left( {p_{{n}}^{\rm{m}} - \chi _{{n}}^{\rm{m}}{q_{\rm{m}}}} \right)}^2} + {{\left( {q_{{n}}^{\rm{m}}} \right)}^2}}]}  \approx \sum\nolimits_{{n}} {{\omega _{{n}}}[ {{{\left( {p_{{n}}^{\rm{m}}} \right)}^2} + {{\left( {q_{{n}}^{\rm{m}} - \chi _{{n}}^{\rm{m}}{p_{\rm{m}}}} \right)}^2}} ]} $.

\section{Details of the derivation of Eqs.~(\ref{eq-4})-(\ref{eq-6})} \label{appendix-B}
In this Appendix, we derive the nonlinear Langevin equations that the total Hamiltonian ${H_{\rm{T}}}$ in Eq.~(\ref{eq-10}) satisfies. To begin with, let us derive the nonlinear Langevin equations satisfied by the optical cavity field.
The Heisenberg equations of motion for the operator $a$ of the optical cavity field and its corresponding reservoir operators ${\Gamma _{{k}}}$ are given by
\begin{eqnarray}
\dot a &=& \frac{1}{{i\hbar }}\left[ {a,{H_{\rm{T}}}} \right] =  - i{\Delta _0}a + i{G_0}a\frac{{\left( {{b^\dag } + b} \right)}}{{\sqrt 2 }} + E - i\sum\limits_{{k}} {{g_{{k}}}} {\Gamma _{{k}}}, \label{eq-B1}\\
{{\dot \Gamma }_{{k}}} &=& \frac{1}{{i\hbar }}\left[ {{\Gamma _{{k}}},{H_{\rm{T}}}} \right] =  - i{\omega _{{k}}}{\Gamma _{{k}}} - i{g_{{k}}}a. \label{eq-B2}
\end{eqnarray}
We are interested in a closed equation for $a$. Equation~(\ref{eq-B2}) for ${{\Gamma _{{k}}}}$ can be formally integrated to yield
\begin{eqnarray}
{\Gamma _{{k}}}\left( t \right) = {\Gamma _{{k}}}\left( {{t_0}} \right){e^{ - i{\omega _{{k}}}\left( {t - {t_0}} \right)}} - i{g_{{k}}}\int_{{t_0}}^t {a\left( \tau  \right)} {e^{ - i{\omega _{{k}}}\left( {t - \tau } \right)}}d\tau.\label{eq-B3}
\end{eqnarray}
Here the first term describes the free evolution of the reservoir modes, whereas the second term arises from their interaction with the optical cavity field. We eliminate ${\Gamma _{{k}}}$ by substituting Eq.~(\ref{eq-B3}) into Eq.~(\ref{eq-B1}), finding
\begin{eqnarray}
\dot a =  - i{\Delta _0}a + i{G_0}a\frac{{\left( {{b^\dag } + b} \right)}}{{\sqrt 2 }} + E - \sum\limits_{{k}} {{{\left( {{g_{{k}}}} \right)}^2}} \int_{{t_0}}^t {a\left( \tau  \right)} {e^{ - i{\omega _{{k}}}\left( {t - \tau } \right)}}d\tau  + {f_a}\left( t \right) \label{eq-B4}
\end{eqnarray}
with ${f_a}\left( t \right) =  - i\sum\nolimits_{{k}} {{g_{{k}}}{\Gamma _{{k}}}\left( {{t_0}} \right)\exp \left[ { - i{\omega _{{k}}}\left( {t - {t_0}} \right)} \right]}$. In Eq.~(\ref{eq-B4}), we can see that the evolution of the system operator depends on the fluctuations in the reservoir. 

To proceed, we introduce some approximations. Following the Weisskopf-Wigner approximation \cite{ref-46}, we replace the summation over $k$ in Eq.~(\ref{eq-B4}) with an integral term, thereby transitioning from a discrete distribution of modes to a continuous one, $\sum\nolimits_{{k}}  \mapsto  {\left( {{L \mathord{\left/{\vphantom {L {2\pi }}} \right.\kern-\nulldelimiterspace} {2\pi }}} \right)^3}\int {{d^3}} k$, where $L$ is the length of the sides of the assumed cubic cavity with no specific boundaries, and ${\rm{\mathord{\buildrel{\lower3pt\hbox{$\scriptscriptstyle\rightharpoonup$}}\over k} }} \equiv \left( {{k_x},{k_y},{k_z}} \right)$ is the wave vector. 

The density of modes between the frequencies $\omega $ and $\omega  + d\omega $ can be obtained by transferring from the Cartesian coordinate to the polar coordinate as in ${\rm{\mathord{\buildrel{\lower3pt\hbox{$\scriptscriptstyle\rightharpoonup$}}\over k} }} \equiv \left( {{k_x},{k_y},{k_z}} \right) \mapsto \left[ {k\sin \left( \theta  \right)\cos \left( \phi  \right),k\sin \left( \theta  \right)\sin \left( \phi  \right),k\cos \left( \theta  \right)} \right]$. The corresponding volume element in the ${{\rm{\mathord{\buildrel{\lower3pt\hbox{$\scriptscriptstyle\rightharpoonup$}}\over k} }}}$ space is ${d^3}k = {k^2}\sin \left( \theta  \right)dkd\theta d\phi  = \left( {{{{\omega ^2}} \mathord{\left/{\vphantom {{{\omega ^2}} {{c^3}}}} \right.\kern-\nulldelimiterspace} {{c^3}}}} \right)\sin \left( \theta  \right)d\omega d\theta d\phi$. The total number of modes ${N_a}$ in the range between $\omega $ and $\omega  + d\omega $ is given by $d{N_a} = {\left( {{L \mathord{\left/{\vphantom {L {2\pi c}}} \right.\kern-\nulldelimiterspace} {2\pi c}}} \right)^3}{\omega ^2}d\omega \int_0^\pi  {\sin \left( \theta  \right)} d\theta \int_0^{2\pi } {d\phi  = \left( {{{{L^3}{\omega ^2}} \mathord{\left/{\vphantom {{{L^3}{\omega ^2}} {2{\pi ^2}{c^3}}}} \right.\kern-\nulldelimiterspace} {2{\pi ^2}{c^3}}}} \right)} d\omega $. A mode density parameter at frequency $\omega $ is therefore given by ${D_a}\left( \omega  \right) = {{d{N_a}\left( \omega  \right)} \mathord{\left/{\vphantom {{d{N_a}\left( \omega  \right)} {d\omega  = }}} \right.\kern-\nulldelimiterspace} {d\omega  = }}{{{L^3}{\omega ^2}} \mathord{\left/{\vphantom {{{L^3}{\omega ^2}} {2{\pi ^2}{c^3}}}} \right.\kern-\nulldelimiterspace} {2{\pi ^2}{c^3}}}$, and ${g_{{k}}} = g\left[ {{\rm{k}}\left( \omega  \right)} \right] = g\left( \omega  \right)$ is the coupling constant evaluated at ${\rm{k}} = {\omega \mathord{\left/{\vphantom {\omega  c}} \right.\kern-\nulldelimiterspace} c}$. We then approximate this spectrum by a continuous spectrum. Thus, the summation in Eq.~(\ref{eq-B4}) can be written as
\begin{eqnarray}
\dot a =  - i{\Delta _0}a + i{G_0}a\frac{{\left( {{b^\dag } + b} \right)}}{{\sqrt 2 }}  + E  - \int_{{t_0}}^t {\int_0^{ + \infty } {{g^2}\left( \omega  \right)} {D_a}\left( \omega  \right){e^{ - i\omega \left( {t - \tau } \right)}}a\left( \tau  \right)d\omega } d\tau + {{ f}_a}\left( t \right). \label{eq-B5}
\end{eqnarray}

Considering an ideal situation, we assume for simplicity that ${\left[ {g\left( \omega  \right)} \right]^2}{D_a}\left( \omega  \right) = {\kappa  \mathord{\left/{\vphantom {\kappa  \pi }} \right.\kern-\nulldelimiterspace} \pi } > 0$ is constant, so that Eq.~(\ref{eq-B5}) is reduced to a simple first-order differential equation \cite{ref-47}:
\begin{eqnarray}
 \dot a =  - i{\Delta _0}a + i{G_0}a\frac{{\left( {{b^\dag } + b} \right)}}{{\sqrt 2 }}  + E - \frac{{{\kappa}}}{\pi }\int_{{t_0}}^{t + {0^ + }} {\int_0^{ + \infty } {{e^{ - i\omega \left( {t - \tau } \right)}}a\left( \tau  \right)d\omega } } d\tau + {{ f}_a}\left( t \right). \label{eq-B6}
\end{eqnarray}
Using the relations
\begin{eqnarray}
\int_0^{ + \infty } {{e^{ - i\omega \left( {t - \tau } \right)}}d\omega }  = {\pi }\delta \left( {t - \tau } \right),\label{eq-B7}
\end{eqnarray}
we arrive at Eq.~(\ref{eq-6}) in the main text:
\begin{eqnarray}
\dot a =  - \left( {{\kappa} + i{\Delta _0}} \right)a + i{G_0}a\frac{{\left( {{b^\dag } + b} \right)}}{{\sqrt 2 }}  + E + \sqrt {2{\kappa}} {a_{{\rm{in}}}}\label{eq-B8}
 \end{eqnarray}
with
\begin{eqnarray}
{a_{{\rm{in}}}}\left( t \right) = \frac{{{f_a}\left( t \right)}}{{\sqrt {2\kappa } }} = \frac{{ - i}}{{\sqrt {2\pi } }}\sum\limits_{{k}} {{g_{{k}}}} \Gamma \left( {{t_0}} \right){e^{ - i{\omega _{{k}}}\left( {t - {t_0}} \right)}}, \label{eq-B9}
 \end{eqnarray}
where ${a_{{\rm{in}}}}\left( t \right)$ is a noise operator which depends upon the environment operators $\Gamma \left( {{t_0}} \right)$ at the initial time, and ${{\kappa}}$ is the decay rate of the optical cavity field, which depends on the coupling strength ${{g_{{k}}}}$ of the optical cavity field and its corresponding reservoirs. We have $q = {{\left( {{b^\dag } + b} \right)} \mathord{/{\vphantom {{\left( {{b^\dag } + b} \right)} {\sqrt 2 }}} \kern-\nulldelimiterspace} {\sqrt 2 }}$ (quadrature definition), and thus we obtain
\begin{eqnarray}
\dot a =  - \left( {{\kappa} + i{\Delta _0}} \right)a + i{G_0}aq + E + \sqrt {2{\kappa}} {a_{{\rm{in}}}}. \label{eq-B10}
\end{eqnarray}

Similarly, the Heisenberg equations of motion for the mechanical operator $b$ and it is corresponding reservoir operators ${\Lambda _{{n}}}$ are given by
\begin{eqnarray}
\dot b &=& \frac{1}{{i\hbar }}\left[ {b,{H_{\rm{T}}}} \right] =  - i{\omega _{\rm{m}}}b + i\frac{{{G_0}}}{{\sqrt 2 }}{a^\dag }a - \frac{1}{{ 2 }}\sum\limits_{{n}} {{\sigma _{{n}}}\left( {\Lambda _{{n}}^\dag  - {\Lambda _{{n}}}} \right)} \label{eq-B11} \\
{{\dot b}^\dag } &=& \frac{1}{{i\hbar }}\left[ {{b^\dag },{H_{\rm{T}}}} \right] = i{\omega _{\rm{m}}}{b^\dag } - i\frac{{{G_0}}}{{\sqrt 2 }}{a^\dag }a + \frac{1}{{ 2 }}\sum\limits_{{n}} {{\sigma _{{n}}}\left( {\Lambda _{{n}}^\dag  - {\Lambda _{{n}}}} \right)} \label{eq-B12}\\
{{\dot \Lambda }_{{n}}} &=& \frac{1}{{i\hbar }}\left[ {{\Lambda _{{n}}},{H_{\rm{T}}}} \right] =  - i{\omega _{{n}}}{\Lambda _{{n}}} - {\sigma _{{n}}}\frac{{\left( {{b^\dag } + b} \right)}}{{ 2 }} \label{eq-B13}\\
\dot \Lambda _{{n}}^\dag  &=& \frac{1}{{i\hbar }}\left[ {\Lambda _{{n}}^\dag ,{H_{\rm{T}}}} \right] = i{\omega _{{n}}}{\Lambda _{{n}}} - {\sigma _{{n}}}\frac{{\left( {{b^\dag } + b} \right)}}{{ 2 }}. \label{eq-B14}
\end{eqnarray}
Since we have the orthogonal relationship $q = {{\left( {{b^\dag } + b} \right)} \mathord{\left/{\vphantom {{\left( {{b^\dag } + b} \right)} {\sqrt 2 }}} \right.\kern-\nulldelimiterspace} {\sqrt 2 }}$ and $p = {{i\left( {{b^\dag } - b} \right)} \mathord{\left/{\vphantom {{i\left( {{b^\dag } - b} \right)} {\sqrt 2 }}} \right.\kern-\nulldelimiterspace} {\sqrt 2 }}$, where $p$ and $q$ are the dimensionless position and momentum operators of the mirror that satisfy the commutation relation $\left[ {q,p} \right] = i$. The derivatives of $q$ and $p$ with respect to time read
\begin{eqnarray}
\dot q &=& \frac{1}{{\sqrt 2 }}\left( {{{\dot b}^\dag } + \dot b} \right) = {\omega _{\rm{m}}}p, \label{eq-B15}\\
\dot p &=& \frac{i}{{\sqrt 2 }}\left( {{{\dot b}^\dag } - \dot b} \right) =  - {\omega _{\rm{m}}}q + {G_0}{a^\dag }a + i\sum\limits_{{n}} {{\sigma _{{n}}}\frac{{\left( {\Lambda _{{n}}^\dag  - {\Lambda _{{n}}}} \right)}}{{\sqrt 2 }}}.\label{eq-B16}
\end{eqnarray}
Equation~(\ref{eq-B15}) corresponds to Eq.~(\ref{eq-4}) in the main text. 

We now focus on a closed equation for $p$. Equations~(\ref{eq-B13})-(\ref{eq-B14}) for ${{\Lambda _{\rm{n}}}}$ and ${\Lambda _{\rm{n}}^\dag }$ can be formally integrated to yield
\begin{eqnarray}
{\Lambda _{{n}}}\left( t \right) &=& {\Lambda _{{n}}}\left( {{t_0}} \right){e^{ - i{\omega _{{n}}}\left( {t - {t_0}} \right)}} - \frac{1}{{ 2 }}{\sigma _{{n}}}\int_{{t_0}}^t {\left[ {{b^\dag }\left( \tau  \right) + b\left( \tau  \right)} \right]} {e^{ - i{\omega _{\rm{n}}}\left( {t - \tau } \right)}}d\tau, \label{eq-B17}\\
\Lambda _{{n}}^\dag \left( t \right) &=& \Lambda _{{n}}^\dag \left( {{t_0}} \right){e^{i{\omega _{{n}}}\left( {t - {t_0}} \right)}} - \frac{1}{{ 2 }}{\sigma _{{n}}}\int_{{t_0}}^t {\left[ {{b^\dag }\left( \tau  \right) + b\left( \tau  \right)} \right]} {e^{i{\omega _{{n}}}\left( {t - \tau } \right)}}d\tau. \label{eq-B18}
\end{eqnarray}
We then eliminate the reservoir operators ${\Lambda _{{n}}}$ and $\Lambda _{{n}}^\dag $ by substituting Eqs.~(\ref{eq-B17})-(\ref{eq-B18}) into Eq.~(\ref{eq-B16}), and obtain
\begin{eqnarray}
\dot p =  - {\omega _{\rm{m}}}q + {G_0}{a^\dag }a + \Theta  + \xi,  \label{eq-B19}
 \end{eqnarray}
where
\begin{eqnarray}
\xi \left( t \right) = \frac{i}{{\sqrt 2 }}\sum\limits_{{n}} {{\sigma _{{n}}}} \left[ {\Lambda _{{n}}^\dag \left( {{t_0}} \right){e^{i{\omega _{{n}}}\left( {t - {t_0}} \right)}} - {\Lambda _{{n}}}\left( {{t_0}} \right){e^{ - i{\omega _{{n}}}\left( {t - {t_0}} \right)}}} \right] \label{eq-B20}
 \end{eqnarray}
and
\begin{eqnarray}
\Theta \left( t \right) = {\sum\limits_{{n}} {\left( {{\sigma _{{n}}}} \right)} ^2}\int_{{t_0}}^t {q\left( \tau  \right)} \sin \left[ {{\omega _{{n}}}\left( {t - \tau } \right)} \right]d\tau.\label{eq-B21}
\end{eqnarray}
Equation~(\ref{eq-B20}) is the same as Eq.~(\ref{eq-8}) in the main text \cite{ref-50,ref-91}. 

We then integrate Eq.~(\ref{eq-B21}) by parts and obtain
\begin{eqnarray}
\Theta \left( t \right) = \sum\limits_{{n}} {\frac{{{{\left( {{\sigma _{{n}}}} \right)}^2}}}{{{\omega _{{n}}}}}} \left\{ {q\left( t \right)\cos \left[ {{\omega _{{n}}}\left( {t - {t_0}} \right)} \right]} \right\}_{{t_0}}^t - \sum\limits_{{n}} {\frac{{{{\left( {{\sigma _{{n}}}} \right)}^2}}}{{{\omega _{{n}}}}}} \int_{{t_0}}^t {\dot q\left( \tau  \right)} \cos \left[ {{\omega _{{n}}}\left( {t - \tau } \right)} \right]d\tau. \label{eq-B22}
\end{eqnarray}
The integrand function $\varsigma \left( t \right) = \sum\nolimits_{{n}} {{{[ {{{\left( {{\sigma _{{n}}}} \right)}^2}\cos \left( {{\omega _{{n}}}t} \right)} ]} \mathord{/{\vphantom {{\left[ {{{\left( {{\sigma _{{n}}}} \right)}^2}\cos \left( {{\omega _{{n}}}t} \right)} \right]} {{\omega _{{n}}}}}} \kern-\nulldelimiterspace} {{\omega _{{n}}}}}} $ can be seen to have the form of memory function since it makes the equation of motion at time $t$ depend on the values of $\dot q\left( t \right)$ for the previous time. Within the Born-Markov approximation \cite{ref-92}, we consider that $\varsigma \left( t \right)$ is a rapidly decaying function and that the equation has a short memory. More precisely, if $\varsigma \left( t \right)$ goes to zero in a time scale that is much less than the time over which $\dot q\left( t \right)$ changes, then we can replace $\dot q\left( \tau  \right)$ by $\dot q\left( t \right)$. For $t$ not close to the initial time ${t_0}$, we can drop the first term in Eq.~(\ref{eq-B22}). Thus, Eq.~(\ref{eq-B22}) reads
\begin{eqnarray}
\Theta \left( t \right) \approx  - \sum\limits_{{n}} {\frac{{{{\left( {{\sigma _{{n}}}} \right)}^2}}}{{{\omega _{{n}}}}}} \int_{{t_0}}^t {\dot q\left( t  \right)} \cos \left[ {{\omega _{{n}}}\left( {t - \tau } \right)} \right]d\tau. \label{eq-B23}
\end{eqnarray}

Similarly to the optical cavity mode $a$, using the Weisskopf-Winger approximation, we consider the spectrum to be given by the normal modes of a large scale, $L \to + \infty$. A difference between phonons and photons is that ${g_{{k}}} = g\left[ {{\rm{k}}\left( \omega  \right)} \right] = g\left( \omega  \right)$ is the coupling constant evaluated at $\omega  \propto {{\rm{k}}^2}$. We then approximate this spectrum by a continuous spectrum. Thus, the summation in Eq.~(\ref{eq-B23}) can be written as
\begin{eqnarray}
\Theta \left( t \right) \approx  - \int_0^{ + \infty } {\int_{{t_0}}^t {d\omega d\tau \frac{{{{\left[ {\sigma \left( \omega  \right)} \right]}^2}}}{\omega }\dot q\left( t \right)} } \cos \left[ {\omega \left( {t - \tau } \right)} \right]{D_b}\left( \omega  \right). \label{eq-B24}
\end{eqnarray}
Considering an ideal situation, by setting ${{{{\left[ {\sigma \left( \omega  \right)} \right]}^2}{D_b}\left( \omega  \right)} \mathord{\left/
 {\vphantom {{{{\left[ {\sigma \left( \omega  \right)} \right]}^2}{D_b}\left( \omega  \right)} \omega }} \right.\kern-\nulldelimiterspace} \omega } = {\gamma  \mathord{\left/{\vphantom {\gamma  \pi }} \right.\kern-\nulldelimiterspace} \pi }$, we thereby obtain
\begin{eqnarray}
\Theta \left( t \right) \approx  - \frac{\gamma }{\pi }\int_0^{ + \infty } {\int_{{t_0}}^{t + {0^ + }} {d\omega d\tau \dot q\left( t \right)} } \cos \left[ {\omega \left( {t - \tau } \right)} \right]. \label{eq-B25}
\end{eqnarray}
Using the relations
\begin{eqnarray}
\int_0^{ + \infty } {\cos \left[ {\omega \left( {t - \tau } \right)} \right]d\omega }  = \pi \delta \left( {t - \tau } \right),\label{eq-B26}
\end{eqnarray}
and by substituting Eq.~(\ref{eq-B15}) into  $\Theta \left( t \right) \approx  - \gamma \dot q\left( t \right)$, we arrive at Eq.~(\ref{eq-5}) in the main text:
\begin{eqnarray}
\dot p =  - {\omega _{\rm{m}}}q - {\gamma _{\rm{m}}}p + {G_0}{a^\dag }a + \xi, \label{eq-B27}
\end{eqnarray}
where the mechanical damping rate is ${\gamma _{\rm{m}}} = {\omega _{\rm{m}}}\gamma$, which depends on the coupling strength ${{\sigma _{{n}}}}$ and the  characteristic frequency of mechanical oscillator ${\omega _{\rm{m}}}$.

\section{Details of the derivation of Eqs.~(\ref{eq-17})-(\ref{eq-18})} \label{appendix-C}
In this appendix, we focus on deriving the nonlinear Langevin equations satisfied by the filtering model~(\ref{eq-11}) under the dominance of resonance effects. Specifically, we concentrate on the mechanical mode $b$, keeping the optical cavity mode $a$ take the same form as the dynamical Eq.~(\ref{eq-6}). By substituting filtering model~(\ref{eq-11}) into the Heisenberg equation, we obtain
\begin{eqnarray}
\dot b &=& \frac{1}{{i\hbar }}\left[ {b,{H_{\rm{T}}}} \right] =  - i{\omega _{\rm{m}}}b + i\frac{{{G_0}}}{{\sqrt 2 }}{a^\dag }a + \frac{1}{2}\sum\limits_{{n}} {{\sigma _{{n}}}{\Lambda _{{n}}}},\label{eq-C1} \\
{{\dot b}^\dag } &=& \frac{1}{{i\hbar }}\left[ {{b^\dag },{H_{\rm{T}}}} \right] = i{\omega _{\rm{m}}}{b^\dag } - i\frac{{{G_0}}}{{\sqrt 2 }}{a^\dag }a + \frac{1}{{ 2 }}\sum\limits_{{n}} {{\sigma _{{n}}}\Lambda _{{n}}^\dag }, \label{eq-C2}\\
{{\dot \Lambda }_{{n}}} &=& \frac{1}{{i\hbar }}\left[ {{\Lambda _{{n}}},{H_{{\rm{F}}}}} \right] =  - i{\omega _{{n}}}{\Lambda _{{n}}} - \frac{{{\sigma _{{n}}}}}{2}b, \label{eq-C3}\\
\dot \Lambda _{{n}}^\dag  &=& \frac{1}{{i\hbar }}\left[ {\Lambda _{{n}}^\dag ,{H_{{\rm{F}}}}} \right] = i{\omega _{{n}}}{\Lambda _{{n}}^\dag} - \frac{{{\sigma _{{n}}}}}{2}{b^\dag }.\label{eq-C4}
\end{eqnarray}
The derivatives of $p$ and $q$ with respect to time read
\begin{eqnarray}
\dot q &=& \frac{1}{{\sqrt 2 }}\left( {{{\dot b}^\dag } + \dot b} \right) = {\omega _{\rm{m}}}p + \frac{1}{2}\sum\limits_{{n}} {{\sigma _{{n}}}{q_{{n}}}}, \label{eq-C5}\\
\dot p &=& \frac{i}{{\sqrt 2 }}\left( {{{\dot b}^\dag } - \dot b} \right) =  - {\omega _{\rm{m}}}q + {G_0}{a^\dag }a + \frac{1}{2}\sum\limits_{{n}} {{\sigma _{{n}}}{p_{{n}}}} \label{eq-C6}.
\end{eqnarray}
We are interested in the system operators $p$ and $q$. Equations~(\ref{eq-C3}) and (\ref{eq-C4}) for ${\Lambda _{{n}}}$ and ${\Lambda _{{n}}^\dag}$ can be formally integrated to yield
\begin{eqnarray}
{\Lambda _{{n}}}\left( t \right) &=& {\Lambda _{{n}}}\left( {{t_0}} \right){e^{ - i{\omega _{{n}}}\left( {t - {t_0}} \right)}} - \frac{{{\sigma _{{n}}}}}{2}\int_{{t_0}}^t {b\left( \tau  \right)} {e^{ - i{\omega _{{n}}}\left( {t - \tau } \right)}}d\tau, \label{eq-C7}\\
\Lambda _{{n}}^\dag \left( t \right) &=& \Lambda _{{n}}^\dag \left( {{t_0}} \right){e^{i{\omega _{{n}}}\left( {t - {t_0}} \right)}} - \frac{{{\sigma _{{n}}}}}{2}\int_{{t_0}}^t {{b^\dag }\left( \tau  \right)} {e^{i{\omega _{{n}}}\left( {t - \tau } \right)}}d\tau. \label{eq-C8}
\end{eqnarray}
The parts of Eqs.~(\ref{eq-C5}) and~(\ref{eq-C6}) that contain environmental operators ${q_{{n}}}$ and ${p_{{n}}}$ can be written as
\begin{eqnarray}
\frac{1}{2}\sum\limits_{{n}} {{\sigma _{{n}}}} {q_{{n}}} &=& \frac{1}{2}\sum\limits_{{n}} {{\sigma _{{n}}}} \frac{{\Lambda _{{n}}^\dag  + {\Lambda _{{n}}}}}{{\sqrt 2 }} = \frac{1}{2}\sum\limits_{{n}} {{\sigma _{{n}}}} \frac{1}{{\sqrt 2 }}\left[ {\Lambda _{{n}}^\dag \left( {{t_0}} \right){e^{i{\omega _{{n}}}\left( {t - {t_0}} \right)}} + {\Lambda _{{n}}}\left( {{t_0}} \right){e^{ - i{\omega _{{n}}}\left( {t - {t_0}} \right)}}} \right]\nonumber\\
 &&- {\sum\limits_{{n}} {\left( {\frac{{{\sigma _{{n}}}}}{2}} \right)} ^2}\frac{1}{{\sqrt 2 }}\left[ {\int_{{t_0}}^t {{b^\dag }\left( \tau  \right)} {e^{i{\omega _{{n}}}\left( {t - \tau } \right)}}d\tau  + \int_{{t_0}}^t {b\left( \tau  \right)} {e^{ - i{\omega _{{n}}}\left( {t - \tau } \right)}}d\tau } \right],\label{eq-C9}\\
\frac{1}{2}\sum\limits_{{n}} {{\sigma _{{n}}}{p_{{n}}}}  &=& \frac{1}{2}\sum\limits_{{n}} {{\sigma _{{n}}}} \frac{{i\left( {\Lambda _{{n}}^\dag  - {\Lambda _{{n}}}} \right)}}{{\sqrt 2 }} = \frac{1}{2}\sum\limits_{{n}} {{\sigma _{{n}}}} \frac{i}{{\sqrt 2 }}\left[ {\Lambda _{{n}}^\dag \left( {{t_0}} \right){e^{i{\omega _{{n}}}\left( {t - {t_0}} \right)}} - {\Lambda _{{n}}}\left( {{t_0}} \right){e^{ - i{\omega _{{n}}}\left( {t - {t_0}} \right)}}} \right]\nonumber\\
 &&- {\sum\limits_{{n}} {\left( {\frac{{{\sigma _{{n}}}}}{2}} \right)} ^2}\frac{i}{{\sqrt 2 }}\left[ {\int_{{t_0}}^t {{b^\dag }\left( \tau  \right)} {e^{i{\omega _{{n}}}\left( {t - \tau } \right)}}d\tau  - \int_{{t_0}}^t {b\left( \tau  \right)} {e^{ - i{\omega _{{n}}}\left( {t - \tau } \right)}}d\tau } \right].\label{eq-C10}
\end{eqnarray}
For convenience, we concisely express Eqs.~(\ref{eq-C9}) and~(\ref{eq-C10}) as
\begin{eqnarray}
\frac{1}{2}\sum\limits_{{n}} {{\sigma _{{n}}}} {q_{{n}}} &=& \frac{1}{2}\xi '\left( t \right) - \chi '\left( t \right),\label{eq-C11}\\
\frac{1}{2}\sum\limits_{{n}} {{\sigma _{{n}}}{p_{{n}}}}  &=& \frac{1}{2}\xi \left( t \right) - \chi \left( t \right)\label{eq-C12},
\end{eqnarray}
where
\begin{eqnarray}
\xi '\left( t \right) &=& \sum\limits_{{n}} {{\sigma _{{n}}}} \frac{1}{{\sqrt 2 }}\left[ {\Lambda _{{n}}^\dag \left( {{t_0}} \right){e^{i{\omega _{{n}}}\left( {t - {t_0}} \right)}} + {\Lambda _{{n}}}\left( {{t_0}} \right){e^{ - i{\omega _{{n}}}\left( {t - {t_0}} \right)}}} \right],\label{eq-C13}\\
\xi \left( t \right) &=& \sum\limits_{{n}} {{\sigma _{{n}}}} \frac{i}{{\sqrt 2 }}\left[ {\Lambda _{{n}}^\dag \left( {{t_0}} \right){e^{i{\omega _{{n}}}\left( {t - {t_0}} \right)}} - {\Lambda _{{n}}}\left( {{t_0}} \right){e^{ - i{\omega _{{n}}}\left( {t - {t_0}} \right)}}} \right],\label{eq-C14}\\
\chi '\left( t \right) &=& \sum\limits_{{n}} {{{\left( {\frac{{{\sigma _{{n}}}}}{2}} \right)}^2}\int_{{t_0}}^t {\left\{ {{q}\left( \tau  \right)\cos \left[ {{\omega _{{n}}}\left( {t - \tau } \right)} \right] + {p}\left( \tau  \right)\sin \left[ {{\omega _{{n}}}\left( {t - \tau } \right)} \right]} \right\}} d\tau }, \label{eq-C15}\\
\chi \left( t \right) &=& \sum\limits_{{n}} {{{\left( {\frac{{{\sigma _{{n}}}}}{2}} \right)}^2}\int_{{t_0}}^t {\left\{ {{p}\left( \tau  \right)\cos \left[ {{\omega _{{n}}}\left( {t - \tau } \right)} \right] - {q}\left( \tau  \right)\sin \left[ {{\omega _{{n}}}\left( {t - \tau } \right)} \right]} \right\}} d\tau }.\label{eq-C16}
\end{eqnarray}

Next, we make some approximations. In a similar way to Appendix~\ref{appendix-B}, under the Born-Markov and Weisskopf-Wigner approximations, Eqs.~(\ref{eq-C15}) and~(\ref{eq-C16}) become
\begin{eqnarray}
\chi '\left( t \right) \!\!&=&\!\!\frac{1}{4}\int_0^{ + \infty }\!\! {\int_{{t_0}}^{t + {0^ + }} {\!\!\!\!d\tau d\omega \left\{ {\dot q\left( t \right)\sin \left[ {\omega \left( {t - \tau } \right)} \right] - \dot p\left( t \right)\cos \left[ {\omega \left( {t - \tau } \right)} \right]} \right\}\frac{{{{\left[ {\sigma \left( \omega  \right)} \right]}^2}{D_b}\left( \omega  \right)}}{\omega }} }\label{eq-C17}\!,\\
\chi \left( t \right) \!\!&=&\!\! \frac{1}{4}\int_0^{ + \infty }\!\! {\int_{{t_0}}^{t + {0^ + }} {\!\!\!\!d\tau d\omega \left\{ {\dot p\left( t \right)\sin \left[ {\omega \left( {t - \tau } \right)} \right] + \dot q\left( t \right)\cos \left[ {\omega \left( {t - \tau } \right)} \right]} \right\}\frac{{{{\left[ {\sigma \left( \omega  \right)} \right]}^2}{D_b}\left( \omega  \right)}}{\omega }} }\!.\label{eq-C18}
\end{eqnarray}
Furthermore, we set ${{{{\left[ {\sigma \left( \omega  \right)} \right]}^2}{D_b}( \omega )} \mathord{/{\vphantom {{{{\left[ {\sigma \left( \omega  \right)} \right]}^2}{D_b}\left( \omega  \right)} \omega }} \kern-\nulldelimiterspace} \omega } = {\gamma  \mathord{\left/{\vphantom {\gamma  \pi }} \right.\kern-\nulldelimiterspace} \pi }$. Then, by using the relation $\int_0^{ + \infty } {\cos \left[ {\omega \left( {t - \tau } \right)} \right]d\omega }  = \pi \delta \left( {t - \tau } \right)$ and $\int_0^{ + \infty } {\sin \left[ {\omega \left( {t - \tau } \right)} \right]d\omega  = } 0$, we find $\chi '\left( t \right) = {{ - \gamma \dot p\left( t \right)} \mathord{\left/{\vphantom {{ - \gamma \dot p\left( t \right)} 4}} \right.\kern-\nulldelimiterspace} 4}$ and $\chi \left( t \right) = {{\gamma \dot q\left( t \right)} \mathord{\left/{\vphantom {{\gamma \dot q\left( t \right)} 4}} \right.\kern-\nulldelimiterspace} 4}$. 

Finally, Eqs.~(\ref{eq-C5}) and~(\ref{eq-C6}) can be rewritten as
\begin{eqnarray}
\dot q &=& {\omega _{\rm{m}}}p + \frac{\gamma }{4}\dot p + \frac{1}{2}\xi ',\label{eq-C19}\\
\dot p &=&  - {\omega _{\rm{m}}}q - \frac{\gamma }{4}\dot q + {G_0}{a^\dag }a + \frac{1}{2}\xi \left( t \right).\label{eq-C20}
\end{eqnarray}
Substituting ${\dot p}$ and ${\dot q}$ into Eqs.~(\ref{eq-C19}) and~(\ref{eq-C20}), respectively, we obtain after decoupling
\begin{eqnarray}
\left( {1 + \frac{{{\gamma ^2}}}{{16}}} \right)\dot q &=& {\omega _{\rm{m}}}p - \frac{{{\gamma _{\rm{m}}}}}{4}q + \frac{\gamma }{4}{G_0}{a^\dag }a + \frac{\gamma }{8}\xi  + \frac{1}{2}\xi ',\label{eq-C21}\\
\left( {1 + \frac{{{\gamma ^2}}}{{16}}} \right)\dot p &=&  - {\omega _{\rm{m}}}q - \frac{{{\gamma _{\rm{m}}}}}{4}p + {G_0}{a^\dag }a - \frac{\gamma }{8}\xi ' + \frac{1}{2}\xi,\label{eq-C22}
\end{eqnarray}
where we set the mechanical damping rate as ${\gamma _{\rm{m}}} = {\omega _{\rm{m}}}\gamma $. Under the weak coupling limit $\gamma  \ll 1$, we neglect the small terms in Eqs.~(\ref{eq-C21}) and~(\ref{eq-C22}) that contain quantities of ${\gamma ^2}$ and $\gamma $. We ultimately reproduce the same Eqs.~(\ref{eq-17}) and~(\ref{eq-18}) as those presented in the main text.

\section{Details of the derivation of the Lyapunov equation~(\ref{eq-31})} \label{appendix-D}
This Appendix derives the Lyapunov equation~(\ref{eq-31}). We begin with the definition of the covariance matrix. According to the definition~\cite{ref-93}, any matrix element of the covariance matrix can be expressed as
\begin{eqnarray}
{{{V}}_{{{ij}}}}\left( t \right) = \frac{1}{2}\left\langle {{\mu _{{i}}}\left( t \right){\mu _{{j}}}\left( t \right) + {\mu _{{j}}}\left( t \right){\mu _{{i}}}\left( t \right)} \right\rangle, \label{eq-D1}
\end{eqnarray}
which satisfies the differential equation
\begin{eqnarray}
\frac{{d{{{V}}_{{{ij}}}}\left( t \right)}}{{dt}} = \frac{1}{2}\left\langle {\frac{{d{\mu _{{i}}}\left( t \right)}}{{dt}}{\mu _{{j}}}\left( t \right) + {\mu _{{i}}}\left( t \right)\frac{{d{\mu _{{j}}}\left( t \right)}}{{dt}} + \frac{{d{\mu _{{j}}}\left( t \right)}}{{dt}}{\mu _{{i}}}\left( t \right) + {\mu _{{j}}}\left( t \right)\frac{{d{\mu _{{i}}}\left( t \right)}}{{dt}}} \right\rangle. \label{eq-D2}
\end{eqnarray}
The matrix elements of the differential Eq.~(\ref{eq-D2}) read
\begin{eqnarray}
{{\dot \mu }_{{i}}}\left( t \right) = \sum\limits_{{o}} {{{{A}}_{{{io}}}}{\mu _{{i}}}\left( t \right) + {n_{{i}}}\left( t \right)}. \label{eq-D3}
\end{eqnarray}
Substituting Eq.~(\ref{eq-D3}) into Eq.~(\ref{eq-D2}), we obtain
\begin{eqnarray}
\frac{{d{{{V}}_{{{ij}}}}\left( t \right)}}{{dt}} &=&  + \frac{1}{2}\left\langle {\left[ {\sum\limits_{{o}} {{{{A}}_{{{io}}}}{\mu _{{o}}}\left( t \right) + {n_{{i}}}\left( t \right)} } \right]{\mu _{{j}}}\left( t \right) + {\mu _{{i}}}\left( t \right)\left[ {\sum\limits_{{o}} {{{{A}}_{{{jo}}}}{\mu _{{o}}}\left( t \right) + {n_{{j}}}\left( t \right)} } \right]} \right\rangle \nonumber\\
 &&+ \frac{1}{2}\left\langle {\left[ {\sum\limits_{{o}} {{{{A}}_{{{jo}}}}{\mu _{{o}}}\left( t \right) + {n_{{j}}}\left( t \right)} } \right]{\mu _{{i}}}\left( t \right) + {\mu _{{j}}}\left( t \right)\left[ {\sum\limits_{{o}} {{{{A}}_{{{io}}}}{\mu _{{o}}}\left( t \right) + {n_{{i}}}\left( t \right)} } \right]} \right\rangle \nonumber\\
 &=& \sum\limits_{{o}} {{{{A}}_{{{io}}}}\left( t \right)} {{{V}}_{{{oj}}}}\left( t \right) + \sum\limits_{{o}} {{{{A}}_{{{jo}}}}\left( t \right)} {{{V}}_{{{io}}}}\left( t \right) + {D_{{{ij}}}\left( t \right)},\label{eq-D4}
\end{eqnarray}
where
\begin{eqnarray}
{D_{{{ij}}}}\left( t \right) = \frac{{\left\langle {{n_{{i}}}\left( t \right){\mu _{{j}}}\left( t \right)} \right\rangle  + \left\langle {{\mu _{{i}}}\left( t \right){n_{{j}}}\left( t \right)} \right\rangle  + \left\langle {{n_{{j}}}\left( t \right){\mu _{{i}}}\left( t \right)} \right\rangle  + \left\langle {{\mu _{{j}}}\left( t \right){n_{{i}}}\left( t \right)} \right\rangle }}{2}.\label{eq-D5}
\end{eqnarray}
We then calculate each term in ${D_{{\rm{ij}}}}$. For example, we have
\begin{eqnarray}
\left\langle {{n_{{i}}}\left( t \right){\mu _{{j}}}\left( t \right)} \right\rangle  &=& \left\langle {{n_{{i}}}\left( t \right)\sum\limits_{{o}} {\left[ {{{ M}_{{{jo}}}}\left( {t,{t_0}} \right){\mu _{{o}}}\left( {{t_0}} \right) + \int_{{t_0}}^t {{{ M}_{{{jo}}}}\left( {t,\tau } \right){n_{{o}}}\left( \tau  \right)d\tau } } \right]} } \right\rangle \nonumber\\
 &=& \sum\limits_{{o}} {{{ M}_{{{jo}}}}\left( {t,{t_0}} \right)\left\langle {{n_{{i}}}\left( t \right){\mu _{{o}}}\left( {{t_0}} \right)} \right\rangle }  + \sum\limits_{{o}} {\int_{{t_0}}^t {{{ M}_{{{jo}}}}\left( {t,\tau } \right)\left\langle {{n_{{i}}}\left( t \right){n_{{j}}}\left( \tau  \right)} \right\rangle d\tau } } \nonumber \\
 &=& \sum\limits_{{o}} {\int_{{t_0}}^t {{{ M}_{{{jo}}}}\left( {t,\tau } \right)\left\langle {{n_{{i}}}\left( t \right){n_{{j}}}\left( \tau  \right)} \right\rangle d\tau } },\label{eq-D6}
\end{eqnarray}
where ${ M}\left( t \right) = \exp \left( {{ A}t} \right)$. Similarly, we obtain the other terms in ${D_{{{ij}}}}$, which are
\begin{eqnarray}
\left\langle {{\mu _{\rm{i}}}\left( t \right){n_{\rm{j}}}\left( t \right)} \right\rangle  &=& \sum\limits_{\rm{o}} {\int_{{t_0}}^t {{{\rm{M}}_{{\rm{io}}}}\left( {t,\tau } \right)\left\langle {{n_{\rm{o}}}\left( \tau  \right){n_{\rm{j}}}\left( t \right)} \right\rangle d\tau } },\label{eq-D7}\\
\left\langle {{n_{\rm{j}}}\left( t \right){\mu _{\rm{i}}}\left( t \right)} \right\rangle  &=& \sum\limits_{\rm{o}} {\int_{{t_0}}^t {{{\rm{M}}_{{\rm{io}}}}\left( {t,\tau } \right)\left\langle {{n_{\rm{j}}}\left( t \right){n_{\rm{o}}}\left( \tau  \right)} \right\rangle d\tau } },\label{eq-D8}\\
\left\langle {{\mu _{\rm{j}}}\left( t \right){n_{\rm{i}}}\left( t \right)} \right\rangle  &=& \sum\limits_{\rm{o}} {\int_{{t_0}}^t {{{\rm{M}}_{{\rm{jo}}}}\left( {t,\tau } \right)\left\langle {{n_{\rm{o}}}\left( \tau  \right){n_{\rm{i}}}\left( t \right)} \right\rangle d\tau } }.\label{eq-D9}
\end{eqnarray}
Hence, ${D_{{{ij}}}}$ can be written as
\begin{eqnarray}
{D_{{{ij}}}} = \sum\limits_{{o}} {\int_{{t_0}}^t {{{ M}_{{{jo}}}}\left( {t,\tau } \right)\Phi _{{{io}}}^{\left( 1 \right)}\left( {t,\tau } \right)d\tau } }  + \sum\limits_{{o}} {\int_{{t_0}}^t {{M_{{{io}}}}\left( {t,\tau } \right)\Phi _{{{oj}}}^{\left( 2 \right)}\left( {t,\tau } \right)d\tau } },\label{eq-D10}
\end{eqnarray}
where
\begin{eqnarray}
\Phi _{{{io}}}^{\left( 1 \right)}\left( {t,\tau } \right) &=& \frac{1}{2}\left\langle {{n_{{i}}}\left( t \right){n_{{o}}}\left( \tau  \right) + {n_{{o}}}\left( \tau  \right){n_{{i}}}\left( t \right)} \right\rangle, \label{eq-D11} \\ \Phi _{{{oj}}}^{\left( 2 \right)}\left( {t,\tau } \right) &=& \frac{1}{2}\left\langle {{n_{{o}}}\left( \tau  \right){n_{{j}}}\left( t \right) + {n_{{j}}}\left( t \right){n_{{o}}}\left( \tau  \right)} \right\rangle. \label{eq-D12}
\end{eqnarray}
The transposes of the column vector of noise operators are given by ${n^T}\left( t \right) = \left[ {0.5\xi '\left( t \right),0.5\xi \left( t \right),\sqrt {2\kappa } {X_{{\rm{in}}}}\left( t \right),\sqrt {2\kappa } {Y_{{\rm{in}}}}\left( t \right)} \right]$, we note that the non-zero correlation functions satisfy the following relations:
\begin{eqnarray}
2\left\langle {{X_{{\rm{in}}}}\left( t \right){Y_{{\rm{in}}}}\left( \tau  \right)} \right\rangle  &=&  - 2\left\langle {{Y_{{\rm{in}}}}\left( t \right){X_{{\rm{in}}}}\left( \tau  \right)} \right\rangle  =  - i\delta \left( {t - \tau } \right), \label{eq-D13}\\
2\left\langle {{X_{{\rm{in}}}}\left( t \right){X_{{\rm{in}}}}\left( \tau  \right)} \right\rangle  &=& 2\left\langle {{Y_{{\rm{in}}}}\left( t \right){Y_{{\rm{in}}}}\left( \tau  \right)} \right\rangle  = \left( {2{{{{\bar n}}}_a} + 1} \right)\delta \left( {t - \tau } \right), \label{eq-D14}\\
\left\langle {\xi \left( t \right)\xi \left( \tau  \right) + \xi \left( \tau  \right)\xi \left( t \right)} \right\rangle  &=& \left\langle {\xi '\left( t \right)\xi '\left( \tau  \right) + \xi '\left( \tau  \right)\xi '\left( t \right)} \right\rangle  = 2{\gamma _{\rm{m}}}\left( {2{{\bar n}} + 1} \right)\delta \left( {t - \tau } \right).\label{eq-D15}
\end{eqnarray}
To be concise, we set ${{{{\rm{\bar n}}}_a}}$ equal to zero. Using the relation~(\ref{eq-D13})-(\ref{eq-D15}), we calculate each term of $\Phi _{{{io}}}^{\left( 1 \right)}\left( {t,\tau } \right)$ and $\Phi _{{{oj}}}^{\left( 2 \right)}\left( {t,\tau } \right)$. The result is given by
\begin{eqnarray}
\Phi _{{{io}}}^{\left( 1 \right)} = \left( {\begin{array}{*{20}{c}}
{\Phi _{11}^{\left( 1 \right)}}&{\Phi _{12}^{\left( 1 \right)}}&{\Phi _{13}^{\left( 1 \right)}}&{\Phi _{14}^{\left( 1 \right)}}\\
{\Phi _{21}^{\left( 1 \right)}}&{\Phi _{22}^{\left( 1 \right)}}&{\Phi _{23}^{\left( 1 \right)}}&{\Phi _{24}^{\left( 1 \right)}}\\
{\Phi _{31}^{\left( 1 \right)}}&{\Phi _{32}^{\left( 1 \right)}}&{\Phi _{33}^{\left( 1 \right)}}&{\Phi _{34}^{\left( 1 \right)}}\\
{\Phi _{41}^{\left( 1 \right)}}&{\Phi _{42}^{\left( 1 \right)}}&{\Phi _{43}^{\left( 1 \right)}}&{\Phi _{44}^{\left( 1 \right)}}
\end{array}} \right) = {{{D}}_{{{io}}}}\delta \left( {t - \tau } \right),\label{eq-D16}
\end{eqnarray}
where ${{{D}}_{{{io}}}}{\rm{ = Diag}}\left[ {{{{\gamma _{\rm{m}}}\left( {2{{\bar n}} + 1} \right)} \mathord{\left/{\vphantom {{{\gamma _{\rm{m}}}\left( {2{{\bar n}} + 1} \right)} 4}} \right.\kern-\nulldelimiterspace} 4},{{{\gamma _{\rm{m}}}\left( {2{{\bar n}} + 1} \right)} \mathord{\left/{\vphantom {{{\gamma _{\rm{m}}}\left( {2{{\bar n}} + 1} \right)} 4}} \right.\kern-\nulldelimiterspace} 4},\kappa ,\kappa } \right]$. 

Similarly, we obtain $\Phi _{{{oj}}}^{\left( 2 \right)} = {{{D}}_{{{oj}}}}\delta \left( {t - \tau } \right) = {\rm{Diag}}\left[ {{{{\gamma _{\rm{m}}}\left( {2{{\bar n}} + 1} \right)} \mathord{\left/{\vphantom {{{\gamma _{\rm{m}}}\left( {2{{\bar n}} + 1} \right)} 4}} \right.\kern-\nulldelimiterspace} 4},{{{\gamma _{\rm{m}}}\left( {2{{\bar n}} + 1} \right)} \mathord{\left/{\vphantom {{{\gamma _{\rm{m}}}\left( {2{{\bar n}} + 1} \right)} 4}} \right.\kern-\nulldelimiterspace} 4},\kappa ,\kappa } \right]\delta \left( {t - \tau } \right)$. Therefore, Eq.~(\ref{eq-D10}) can be rewritten as
\begin{eqnarray}
{{{D}}_{{{ij}}}} &=& \sum\limits_{{o}} {\int_{{t_0}}^t {{{ M}_{{{jo}}}}\left( {t,\tau } \right)\Phi _{{{io}}}^{\left( 1 \right)}\left( {t,\tau } \right)d\tau } }  + \sum\limits_{{o}} {\int_0^t {{{ M}_{{{io}}}}\left( {t,\tau } \right)\Phi _{{{oj}}}^{\left( 2 \right)}\left( {t,\tau } \right)d\tau } }\nonumber\\
 &=& \sum\limits_{{o}} {\int_{{t_0}}^t {{{ M}_{{{jo}}}}\left( {t,\tau } \right){{{D}}_{{{io}}}}\delta \left( {t - \tau } \right)d\tau } }  + \sum\limits_{{o}} {\int_{{t_0}}^t {{{ M}_{{{io}}}}\left( {t,\tau } \right){{{D}}_{{{oj}}}}\delta \left( {t - \tau } \right)d\tau } }\nonumber \\
 &=& \frac{1}{2}\sum\limits_{{o}} {{{\rm I}_{{{jo}}}}{{{D}}_{{{io}}}} + } \frac{1}{2}\sum\limits_{{o}} {{{\rm I}_{{{io}}}}{{{D}}_{{{oj}}}}}  = \frac{1}{2}\sum\limits_{{o}} {{{{D}}_{{{io}}}}{\rm{I}}_{{{oj}}}^T + } \frac{1}{2}\sum\limits_{{o}} {{{\rm{I}}_{{{io}}}}{{{D}}_{{{oj}}}}}  \equiv {{D}},\label{D17}
\end{eqnarray}
where ${{D = {\rm{Diag}}}}\left[ {{{{\gamma _{\rm{m}}}\left( {2{{\bar n}} + 1} \right)} \mathord{\left/{\vphantom {{{\gamma _{\rm{m}}}\left( {2{{\bar n}} + 1} \right)} 4}} \right.\kern-\nulldelimiterspace} 4},{{{\gamma _{\rm{m}}}\left( {2{{\bar n}} + 1} \right)} \mathord{\left/{\vphantom {{{\gamma _{\rm{m}}}\left( {2{{\bar n}} + 1} \right)} 4}} \right.\kern-\nulldelimiterspace} 4},\kappa ,\kappa } \right]$.

Thus, Eq.~(\ref{eq-D4}) reads ${{\dot V = }}{ A}{{V + V}}{{ A}^T} + D$. When the stability conditions are satisfied, in the long-time limit, the derivative of the covariance matrix with respect to time approaches zero, ${{\dot V}} = 0$. This produces the Lyapunov equation~(\ref{eq-31}) in the main text, ${{AV}} + {{V}}{{{A}}^T} =  - {{D}}$.
\end{widetext}

\end{document}